\documentclass[twocolumn]{aastex631}

\usepackage{amsmath}
\usepackage{graphicx}

\newlength{\subcolumnwidth}

\newcommand{\nextsubcolumn}[1][]{%
  \cr\noalign{\hfill}
  \if\relax\detokenize{#1}\relax\else\hsize=#1\setlength{\subcolumnwidth}{\hsize}\fi
}

\newcommand{\rrr}[1]{\textcolor{black}{#1}}

\newcommand{\halpha}{H$\alpha$}

\newcommand{\msun}{$M_{\odot}$}

\newcommand{\kms}{\text{ km s}^{-1}}
\NewDocumentCommand{\per}{O{1} m}{^{-#2}}
\newcommand{\rvir}{$R_{\rm vir}$}

\newcommand{\bbarolo}{\textsf{\textsuperscript{3D}Barolo}}

\newcommand{\tinytitans}{TiNy Titans}

\newcommand{\companion}{PGC027864}
\newcommand{\bigneighbor}{IC0560}
\newcommand{\ugc}{UGC5205}
\newcommand{\lurker}{NSA6}
\newcommand{\acquaintance}{NSA6}
\newcommand{\sdssvelocity}{$1500\pm 18\kms{}$}

\newcommand{\HI}{\ion{H}{1}}

\newcommand{\HH}{${\rm H_2}$}

\newcommand{\ionline}[3]{[\ion{#1}{#2}]$\lambda$#3\AA{}}

\shorttitle{\ugc{}}
\shortauthors{Kado-Fong et al.}

\graphicspath{{./}{FIG/}}

\begin{document}

\title{Dwarf-Dwarf Interactions Can Both Trigger and Quench Star Formation}

\author[0000-0002-0332-177X]{Erin Kado-Fong}
\affiliation{Physics Department, Yale Center for Astronomy \& Astrophysics, PO Box 208120, New Haven, CT 06520, USA}
\correspondingauthor{Erin Kado-Fong} 
\email{erin.kado-fong@yale.edu}
\author{Azia Robinson}
\affiliation{Department of Astrophysical Sciences, Princeton University, Princeton, NJ 08540, USA}
\author[0000-0003-1991-370X]{Kristina Nyland}
\affiliation{U.S. Naval Research Laboratory, 4555 Overlook Ave SW, Washington, DC 20375, USA}
\author[0000-0002-5612-3427]{Jenny E. Greene}
\affiliation{Department of Astrophysical Sciences, Princeton University, Princeton, NJ 08540, USA}
\author[0000-0002-1714-1905]{Katherine A. Suess}\altaffiliation{NHFP Hubble Fellow}
\affiliation{Kavli Institute for Particle Astrophysics and Cosmology and Department of Physics, Stanford University, Stanford, CA 94305, USA}
\author[0000-0002-2596-8531]{Sabrina Stierwalt}
\affiliation{Physics Department, 1600 Campus Road, Occidental College, Los Angeles, CA 90041, USA}
\author[0000-0002-1691-8217]{Rachael Beaton}
\affiliation{Space Telescope Science Institute, Baltimore, MD 21218, USA}
\affiliation{Department of Physics and Astronomy, Johns Hopkins University, Baltimore, MD 21218, USA}

\begin{abstract}
It is exceedingly rare to find quiescent low-mass galaxies in the field \rrr{at low redshift}. \ugc{} is an example of such a quenched field dwarf ($M_\star\sim3\times10^8M_\odot$). Despite a wealth of cold gas ($M_{\rm HI}\sim 3.5 \times 10^8 M_\odot$) and \rrr{ultraviolet} emission that 
indicates significant star formation in the past few hundred Myr, there is no detection of \halpha{} 
emission -- star formation in the last $\sim 10$ Myr -- across the face of the galaxy. 
Meanwhile, the near equal-mass companion of \ugc{}, \companion{}, is starbursting ($\rm EW_{\rm H\alpha}>1000$ \AA). 
In this work, we present
new Karl G. Jansky Very Large Array (VLA) 21 cm line observations of \ugc{} \rrr{showing that} the lack of star formation is caused by an absence of \HI{} in the main body of the galaxy. The \HI{} of \ugc{} is highly disturbed; the bulk of the \HI{} resides in several kpc-long tails, while the \HI{} of \companion{} is dominated by ordered rotation. We model the stellar populations of \ugc{} to show that, as indicated by the UV-\halpha{} emission, the galaxy underwent a coordinated 
quenching event $\sim\!100-300$ Myr ago. The asymmetry of outcomes for \ugc{} and \companion{} demonstrate that major mergers 
can both quench \textit{and} trigger star formation in dwarfs. However, because the gas remains bound to the system, we suggest that such mergers only temporarily quench star formation. We estimate a total quenched time of $\sim 560$ Myr for \ugc{}, consistent with established upper limits  on the quenched fraction of a few percent for dwarfs in the field. 
\end{abstract}

\section{Introduction}






\begin{figure*}[t]
  \centering     
  \makebox[\textwidth][c]{\includegraphics[width=1.2\textwidth]{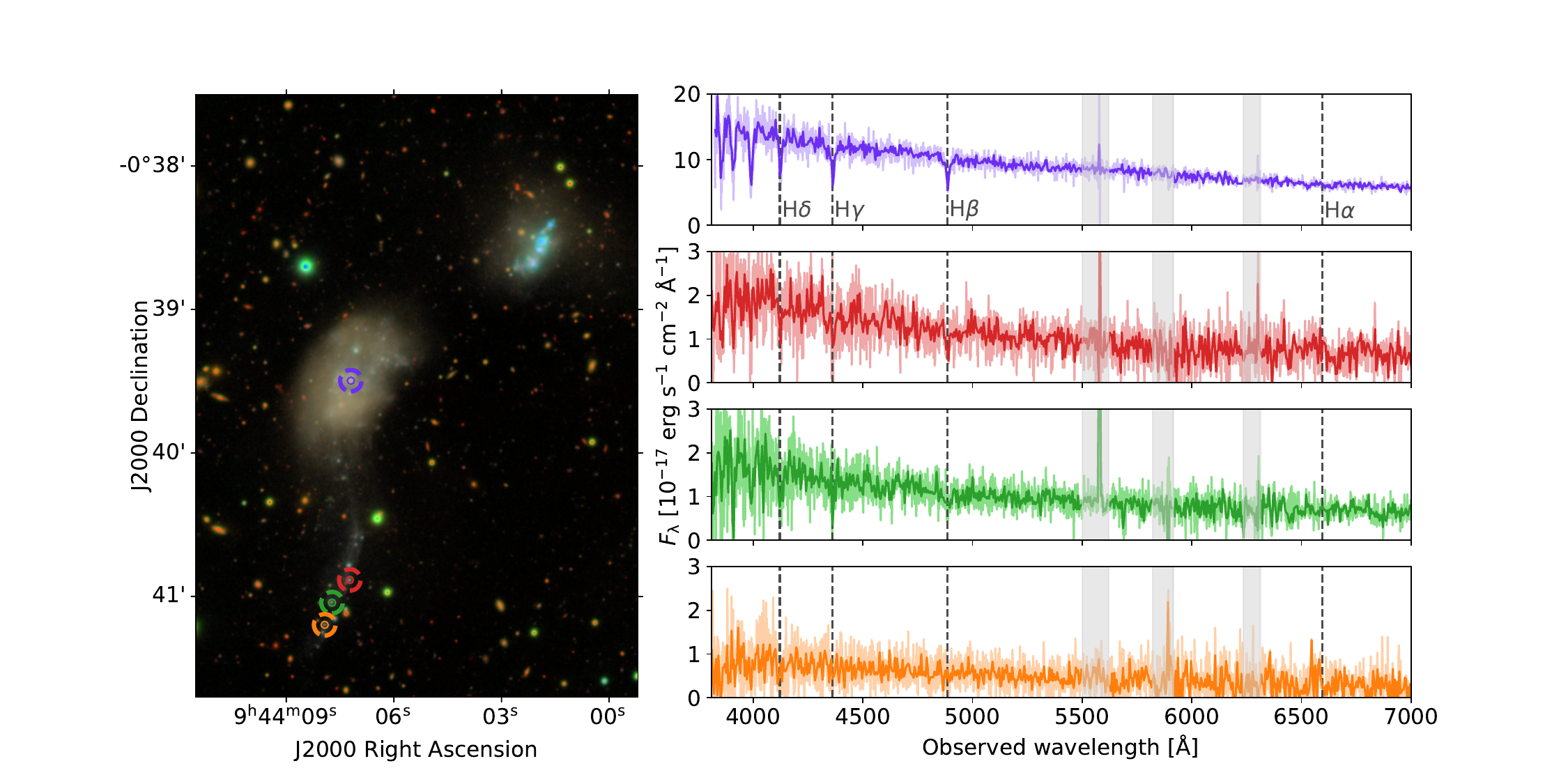}}
  \vspace{-25pt}
  \caption{\textit{Left:} HSC-SSP $gri$ composite image of \ugc{} (southeast) and 
          \companion{} (northwest). The locations of the four archival SDSS spectra
          are shown in orange, green, red, and purple. \rrr{The inner circles show the extent of the SDSS fiber 3'' diameter.} 
          \textit{Right:} the archival SDSS
          spectra (same colors as at left). \rrr{We show the unbinned spectra as light curves and the spectra binned to $\Delta \lambda=5$ \AA{} as bold curves.} Grey shaded patches show regions with
          under-subtracted sky lines. The blue dashed lines show the Balmer series 
          redshifted to $z=0.00496$; there are no significant emission lines in the spectra of 
          \ugc{}.}
  \label{fig:sdsshsc}
\end{figure*}

Dwarf galaxies ($M_* < 10^9$~\msun) provide a crucial laboratory for the processes, both internal and environmental, that regulate star-formation. Unlike more massive galaxies, empirical evidence 
suggests that low-mass galaxies are nearly always star-forming in the field, with less than $1\%$ of galaxies at
$M_\star =10^8 M_\odot$ expected to be quenched outside the influence of a massive galaxy \citep{geha_2012}.
While merging and 
feedback are actively discussed as mechanisms for truncating star formation in massive galaxies \citep[e.g.,][]{Blanton:2009aa}, the dearth of a build-up of quenched dwarfs in the field suggests that these mechanisms either do not exist in the dwarf regime, or are inefficient at quenching dwarf galaxies. 

In this paper, we study an intriguing exception that may prove the rule. \ugc{} is a dwarf galaxy ($M_\star=3\times10^8M_\odot$) \rrr{at $d\approx21$ Mpc}
that is classified as a field dwarf by standard criteria \citep{geha_2012}.
However, \ugc{} is in an 
interacting pair with the comparable-mass galaxy \companion{}. This pair is part of the \tinytitans{} survey of dwarf interacting pairs 
\citep{stierwalt2015,stierwalt2017}.
While both galaxies are very \ion{H}{1} rich, there is no sign of \halpha{} emission across the face of \ugc{} \citep{Pracy:2012}. 
\companion{}, meanwhile, is 
vigorously star-forming, and has been studied in the UV as a local analog to high redshift galaxies \citep[e.g.,][]{izotov2016,senchyna2022}. The two galaxies in this pair thus have radically different effective star formation efficiencies (where we will roughly define star formation efficiency as $\text{SFE}=\text{SFR}/M_{\rm HI}$) despite very similar stellar masses, \HI{} masses, and large scale environments.

Our aim in the present work is to understand the root of this apparent disparity in SFE 
by studying the morphology and kinematics of the \ion{H}{1} gas, particularly in the quenched dwarf \ugc{}. 
While mergers are a recognized mechanism for quenching star formation in massive galaxies \citep[see, e.g.][]{hopkins2008, somerville2008, quai2021}, this pathway has not been explored as a quenching mechanism 
for dwarfs. 
In fact, although
signatures of dwarf-dwarf interactions and mergers at this mass ratio are observed in a few percent of dwarfs \citep{stierwalt2015,besla:2018,kado-fong:2020aa}, for both dwarf pairs \citep{stierwalt2015} and post-coalescence mergers \citep{kado-fong:2020aa}, dwarf-dwarf interactions appear to trigger, rather than quench, star formation. In particular,  long-term quenching via low-mass mergers  is disfavored due to the lack of an emergent field dwarf population at low redshift \citep{geha_2012} despite theoretical expectations that major mergers between dwarfs are relatively common ($\sim 10\%$) since $z\sim 1$ and direct evidence for major mergers at low-$z$ \citep[see, e.g.][]{stierwalt2015, kado-fong:2020aa}. 
\ugc{} thus provides an excellent laboratory to interrogate the physical processes at play during dwarf-dwarf mergers, and how quenching can occur dynamically in such a gas-rich system.

This manuscript is structured as follows. \autoref{s:observations} introduces the {Karl G. Jansky Very Large Array (VLA)} data that we obtained of \ugc{} and \companion{}, while \autoref{s:results} discusses the analysis of the VLA data and archival UV-optical datasets. Finally, in \autoref{s:discussion} we discuss the viability of proposed mechanisms to quench \ugc{} as well as the implications that the existence of \ugc{} has on the existence and evolution of a quiescent dwarf population.
 
\rrr{
Throughout this analysis, we adopt a flat $\Lambda$CDM cosmology with $\Omega_m=0.3$ ($\Omega_\Lambda=0.7$), and
$H_0=70\ \kms{}\ {\rm Mpc^{-1}}$. 
}

\section{Observations and Data Reduction}\label{s:observations}

In this section we present and discuss new {multi-configuration VLA observations from project 21A-081.}  We observed \ugc{} in the L-band centered at 1413.8929 MHz in the D configuration on March 25, 2021 for 2 hr (1.5 hr on source).   
{On June 13 and 15, 2021 we observed \ugc{} with an identical frequency tuning in the C configuration for} 3.5 hr (2.5 hr on source) each, for a total of 5 hr on source.  Phase referencing using the calibrator J0943-0819 was performed every three minutes. Once per observation, we performed flux density  and bandpass calibration using the calibrator 3C138 for five minutes.

Initial flagging, flux density scaling, and  calibration for the three measurement sets were performed with the Common Astronomy Software Applications (CASA; \citealt{2022PASP..134k4501C}) pipeline 
{version 1.4.2\footnote{https://science.nrao.edu/facilities/vla/data-processing/pipeline/scripted-pipeline},} %
with additional flagging, data reduction, and imaging performed with 
CASA version 6.4.3.27. 
We reduced each of the two C configuration scheduling blocks separately; we combined the reduced datasets from each block as the final step to produce the final C configuration dataset.
{We note that the C configuration data contained severe radio frequency interference (RFI) requiring manual flagging using the {\tt PLOTMS} and {\tt FLAGDATA} tasks.}

{For each configuration, we used the task {\tt TCLEAN} to produce an initial image cube containing both the line and continuum emission.  We imaged the data with a spectral resolution of 6.6 km/s and chose `Briggs' weighting with a robust parameter of 0.3 to achieve an optimal compromise among sensitivity, angular resolution, and sidelobe suppression. 
 The cell sizes used for imaging the C and D configuration datasets were 2.2 and 9 arcseconds, respectively.}   
The HI emission associated with \ugc{} and MGC+00-25-010 spanned channels 224 to 252 (1575 km/s to 1395 km/s). 
{We performed continuum subtraction}  
using the CASA task {\tt UVCONTSUB} before 
{ producing the final image cube for each configuration} with {\tt TCLEAN}. 
Our final C configuration image achieved an rms noise of 
$7.6 \, \rm{mJy \, beam^{-1}}$ 
and our final D configuration image reached an rms of 
$1.5 \, \rm{mJy \, beam^{-1}}$.

In the subsequent sections, we will consider the VLA data coincident with 
\ugc{} and with \companion{} separately (in \autoref{s:results:ugc} and 
\autoref{s:results:companion}, respectively). For the reader's convenience,
however, we tabulate key VLA figures here. The C-array moment zero map of the system is shown in \autoref{f:moment_zero}, while the C-array moment one map of the system 
is shown in \autoref{f:moment_one}. The moment zero map of the D-array data 
is shown in \autoref{f:moment_zero_DARRAY}. Channel maps of \ugc{} and
\companion{} are shown separately in \autoref{fig:ChanUGCa} and 
\autoref{fig:ChanUGCc}, respectively.

\section{Results}\label{s:results}

\subsection{Optical Morphology}
Before presenting the spatially resolved \HI{} data obtained with VLA, we will briefly detail the optical morphology of the 
galaxy pair in order to provide context to the reader in interpreting the gas reservoirs of these two galaxies.

In Figure \ref{fig:sdsshsc}, we show an image from the Hyper-Suprime Camera data of the \ugc{}/\companion{} pair in the $gri-$bands. 
Superposed on the image are the locations of the four SDSS fibers that took the spectra shown on the right-hand side. We point out a few key 
features of this interacting system. First, there is clearly a tidal feature stretching to the South of \ugc{}. Likely this tail is 
associated with the ongoing interaction with \companion{} to the North. The striking point, and the reason for our focus on \ugc{}, is that despite the clear evidence of young stellar ages in the 
spectra (a point to which we will return more quantitatively in \autoref{s:results:stellarpops}), 
there is no H$\alpha$ detected at any position covered by the SDSS. Likewise an integral-field observation of the galaxy center shows no H$\alpha$ emission \citep{Pracy:2012}. 

In contrast, as is 
clear from this three-color image, \companion{} is star-bursting. The low metallicity and high star formation rate of this galaxy have been documented extensively \citep[e.g.,][]{Shirazi:2012,Senchyna:2017}. 

\begin{figure}[t]
  \centering
  \includegraphics[width=\linewidth]{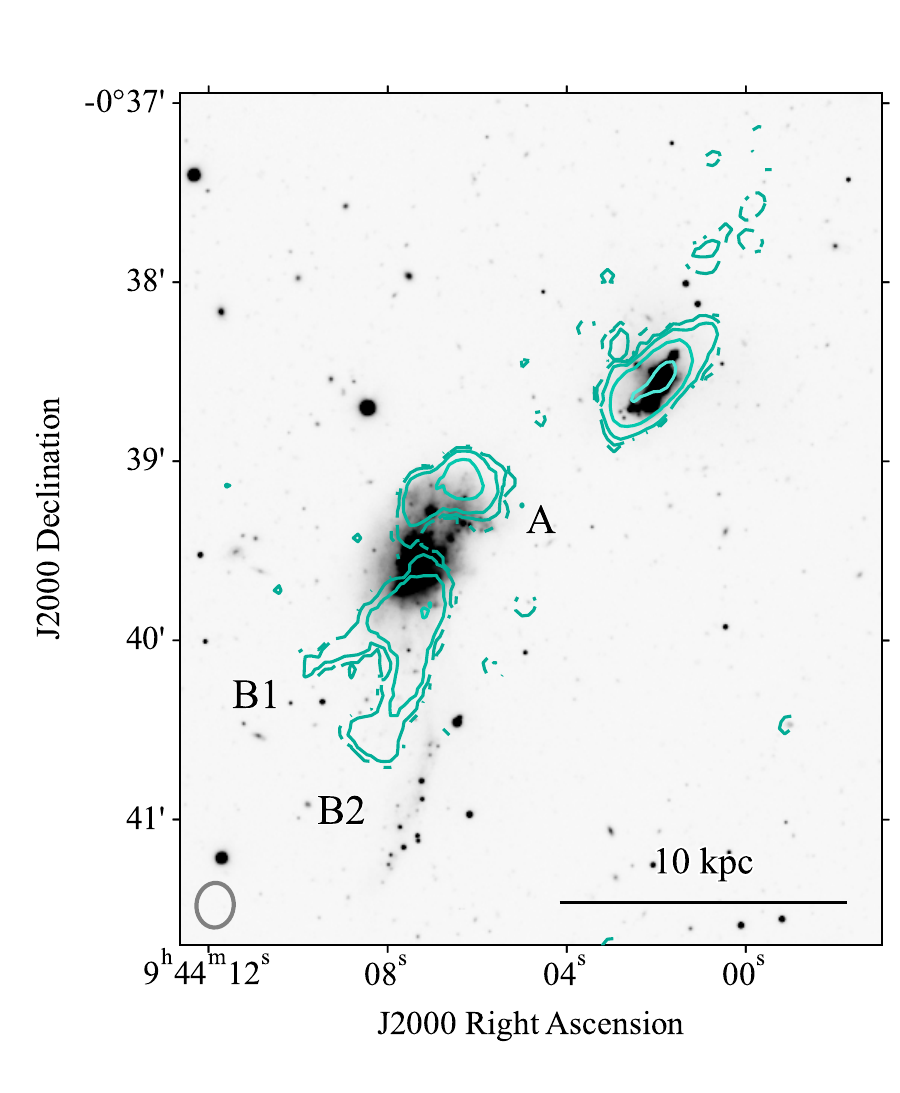}
  \caption{
      The moment 0 map of the VLA C-array data (green contours), overlaid 
      over the HSC-SSP g-band optical image. Significant \HI{} gas is 
      found coincident with both    
      \ugc{} (southeast) and \companion{} (northwest). We find that the 
      gas associated with \ugc{} is morphologically complex, with three 
      clear tails (labels A, B1, and B2) visible in the VLA data.
      The gas associated with \companion{} is notably less disturbed, though 
      there is significant emission from a disturbed component to the 
      northwest. We show contours with a constant logarithmic offset 
      beginning from $\Sigma_{\rm HI} = 10^5 M_\odot$ beam$^{-1}$ (or an \HI{} column density of $N({\rm HI})=5.4\times 10^{18}$ cm$\per{2}$). 
  }\label{f:moment_zero}
  \end{figure}

Having addressed the optical morphology, we can now consider the spatially resolved \HI{} data 
for this galaxy pair. The moment 0 maps of the C-array VLA data are shown as contours overlaid on the HSC-SSP optical imaging of \ugc{} and \companion{} in \autoref{f:moment_zero}.
We see immediately see that the \ion{H}{1} associated with \ugc{} has a complex morphology and the \HI{} associated with \companion{} is relatively less disturbed. 

  \begin{figure*}[t]
  \centering
  \includegraphics[width=\linewidth]{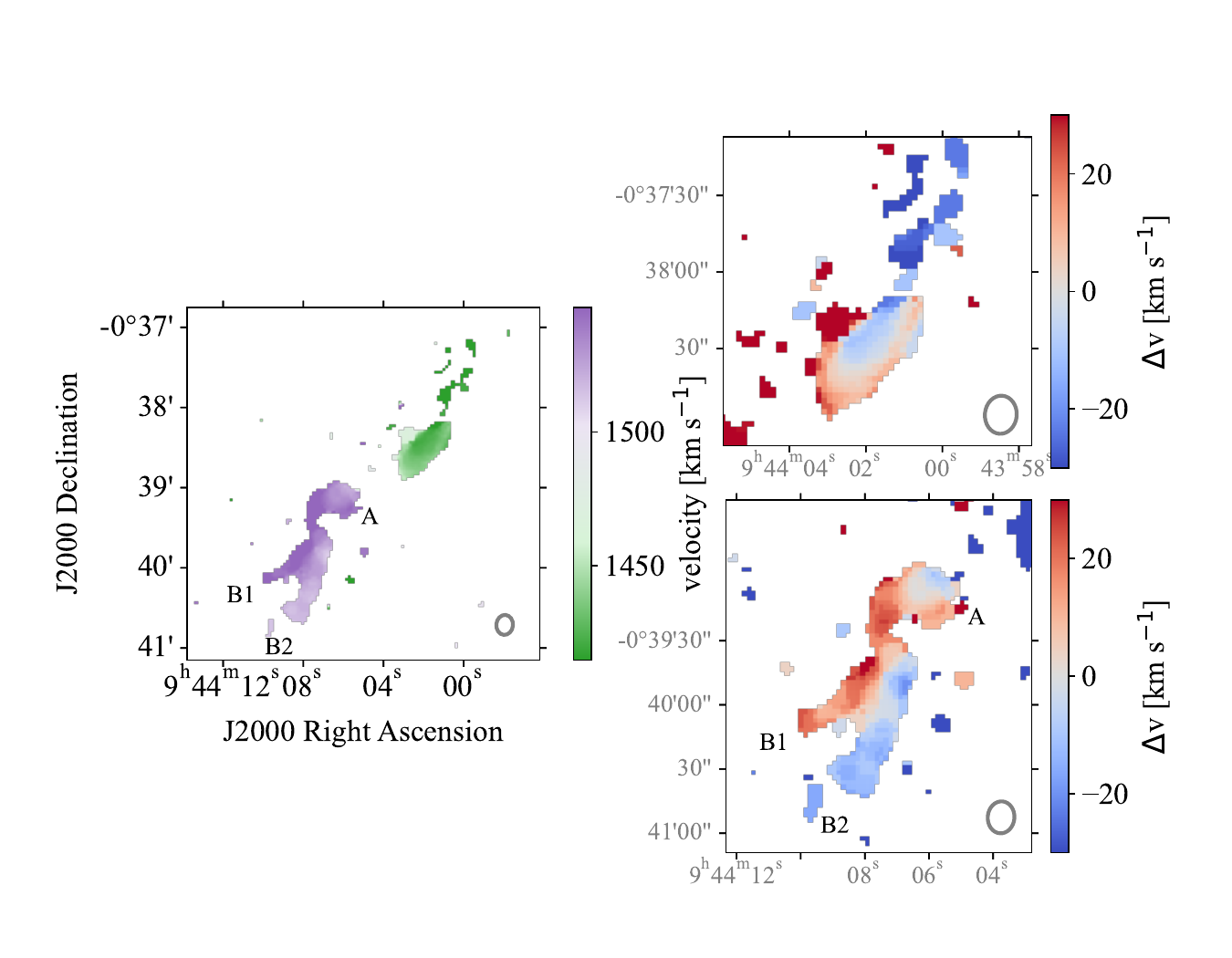}
  \vspace{-30pt}
  \caption{
      The first moment map of \ugc{} and \companion{}. At left we show the 
      full moment map, while at right we show a portion of the map 
      over a smaller spatial and velocity range. In each panel, the 
      average C-array beam is shown in the bottom right corner. 
      We find that the \HI{} associated with \ugc{} and that associated with 
      \companion{} are systematically offset in velocity. The 
      gas associated with \ugc{} is highly disturbed, with extended features 
      A, B1, \& B2 (as labeled) 
      showing evidence of an offset in mean velocity. 
      Though there is some blueshifted emission of the gas associated 
      with \companion{}, the majority of the emission is consistent with
      ordered rotation.
  }\label{f:moment_one}
  \end{figure*}

\subsection{\HI{} morphology and kinematics of \ugc{}}\label{s:results:ugc}
There are three primary components of the \HI{} associated with \ugc{} that we will consider throughout 
the paper, as labeled in \autoref{f:moment_zero}. 
While there appears to be 
gas to the southeast of the galaxy (components B1 \& B2) \rrr{and} gas to the northwest of the galaxy (component A), there is no detected concentration of \ion{H}{1} in the galaxy center. 

To better understand the nature of the complex \HI{} morphology of \ugc{}, we 
show the first moment maps in \autoref{f:moment_one}. Here, the left panel shows the full extent of the 
\HI{} associated with \ugc{} and \companion{}, while the right top and bottom panels
show the difference in velocity from the peak heliocentric velocity of the \HI{} profile
of \ugc{} and \companion{}, respectively.

From the moment 1 maps, we note the varied velocities along B1 and B2. The systemic velocity of the stars, as measured from the absorption features in the SDSS spectra, is \sdssvelocity{}. Nearly all of 
the gas is redshifted relative to this velocity, with only the very tip of tail B2 overlapping in velocity with the galaxy systemic 
velocity. This velocity mismatch was previously established by single-dish spectra (see \autoref{s:appendix:archival}). Tail B1 and component A share a similar velocity of 1500-1550~km~s$^{-1}$, and there is a small gradient across component A. 

\begin{figure*}
    \centering
    \includegraphics[width=\linewidth]{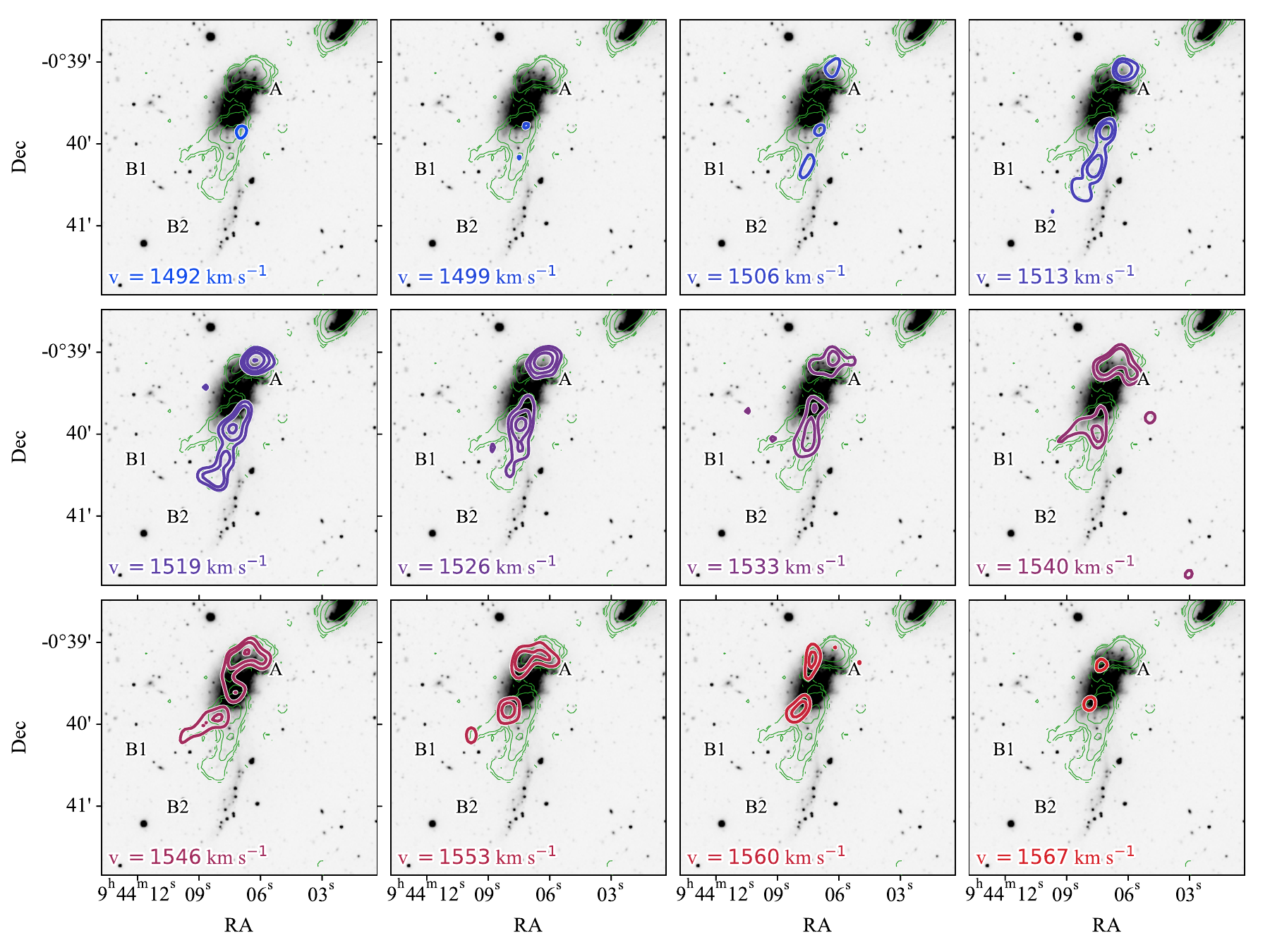}
    \caption{Channel maps for channels where significant 21cm line emission is
    detected around \ugc{}. As a visual guide, the
    tidal components of \ugc{} (A, B1, and B2) labeled and the
    moment 0 \HI{} map are shown as green contours in each panel.
    The narrow width of the \HI{} components A, B1, and B2 in velocity space support the interpretation that these 
    features are tidal in nature.
    The contours of the channel maps begin at 3 times the 
    RMS of the C-array data (0.76 mJy/beam) and increase
    by a factor of $2^n$ at the $n^{\rm th}$ contour.}
\label{fig:ChanUGCa}
\end{figure*}

\rrr{B2} extends away from the galaxy center in a similar direction as the stellar tidal feature. Because of the spatial proximity to the stellar tidal feature, and due to the narrow range of velocities we see in B1 and B2 in the moment 1 map, we suggest that B1 and B2 are tidal in nature.  We show the channel maps of \ugc{} in \autoref{fig:ChanUGCa} -- 
B1 and B2 span only a couple of channels, and we see little to no 
gradient along the features.
Both B1 and B2 are redshifted relative to \ugc{}, but their velocity structure seems quite typical for tidal tails.

\begin{figure*}
  \centering
\gridline{
        \includegraphics[width=0.45\textwidth]{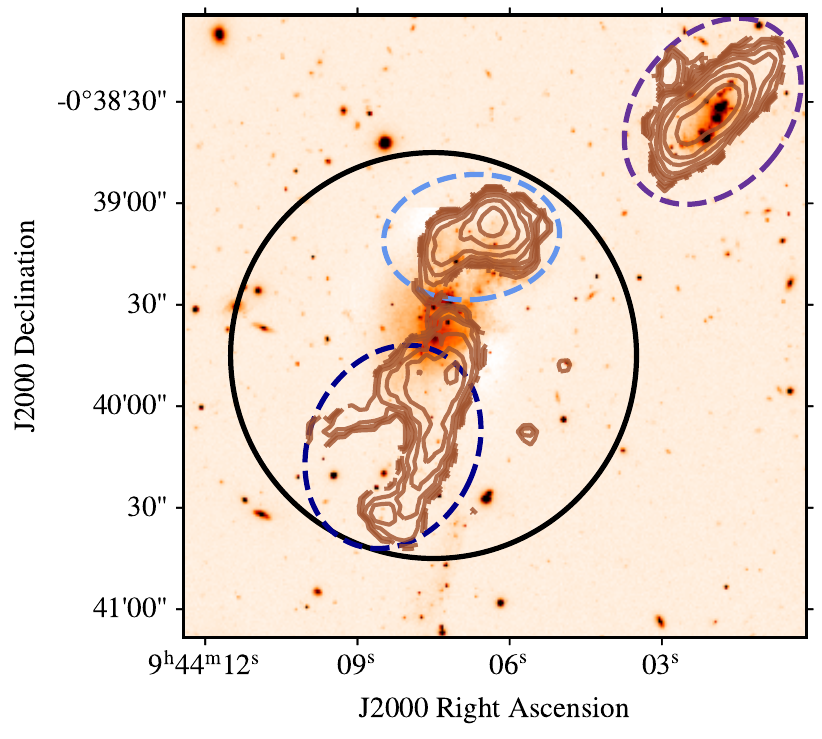}
        \includegraphics[width=0.45\textwidth]{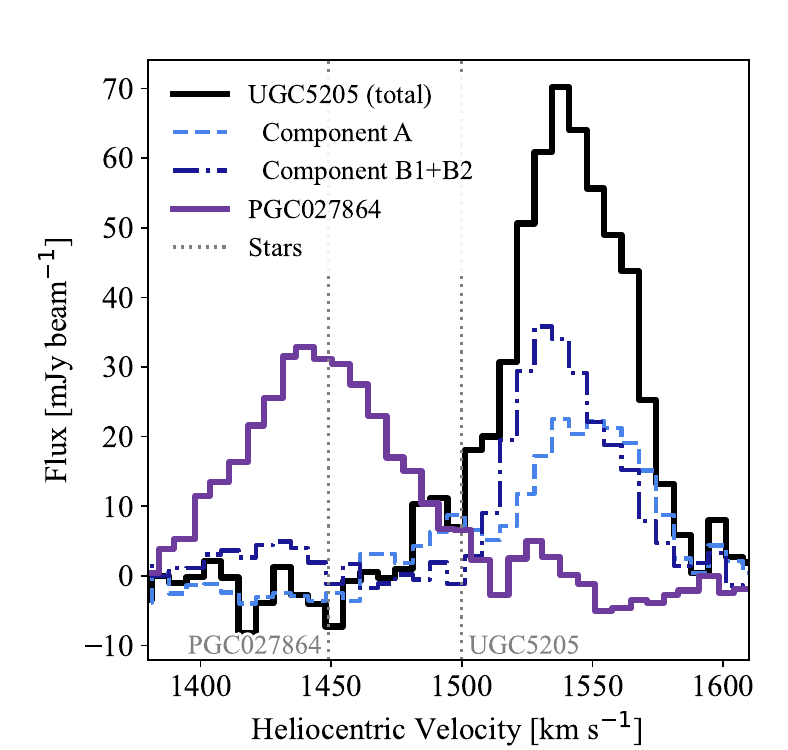}        
        }    
      \caption{Integrated spectra of each compoment of the \HI{} emission associated with \ugc{} and \companion{}. Each ellipse at left indicates an extraction region corresponding to an integrated spectrum at right as indicated by color; 
      note that in this figure, component A is encompassed by 
      the ``head'' extraction region 
      while both components B1 and B2 are within the ``Tail'' extraction region.
s      All spectra have been extracted using CASA IMVIEW. The grey dashed vertical line shows the systemic optical velocity of \ugc{}.}\label{fig:integratedspectra}
\end{figure*}

The nature of Feature A, which sits between the two galaxies spatially but is much closer to the systemic velocity of \ugc{}, is unclear. The total mass of \ion{H}
{1} associated with \ugc{} is $3.47 \times 10^{8} M_{\odot}$. \rrr{The mass in \HI{} associated with the extended features is} split almost equally between A and B1+B2. Specifically, as detailed in \autoref{t:hicomponents}, component A has an \ion{H}{1} mass of $1.33 \times 10^{8} M_{\odot}$ while the two tails have a combined mass of $1.41 \times 10^{8} M_{\odot}$. The apertures used to measure the enclosed \HI{} mass in 
each component are shown as dashed ellipses in \autoref{fig:integratedspectra}.

Component A spans a range of velocities ($1520\lesssim v\lesssim 1570$~km~s$^{-1}$). 
There is a some evidence for a velocity gradient across Component A, with the material to the East falling at higher velocity 
and the material to the West (closest to the companion galaxy) at the lowest velocity, as can be seen directly from the channel maps in \autoref{fig:ChanUGCa}. 
Component A, recall, comprises roughly 50\% of the total 
\ion{H}{1} associated with \ugc{}. 

\begin{deluxetable}{lcccc}
  \tablecaption{Gas Components. \label{tab:components}}
  \tablewidth{0pt}
  \tablecaption{\HI{} masses associated with each component of \ugc{}.}
  \tablehead{
  \colhead{Galaxy} & \colhead{Component} &  \colhead{Velocity} &  \colhead{Channel} & \colhead{$M_{HI}$$^a$} \\
  \colhead{} & \colhead{} & \colhead{km s$^{-1}$} & & \colhead{$10^8\ M_\odot$}
  }
  \startdata
      \ugc{}         &  \textit{total}           & $1575-1507$ & $224\!-\!234$   & 3.47     \\
                      & A           & $1575\!-\!1507$ & \nodata     & 1.33     \\
                      & B1          & $1547\!-\!150$0 & $228\!-\!235$   & 1.41  \\
                      & B2          & $1575\!-\!1547$ & $224\!-\!228$   &  \nodata \\
    \hline{}   
      \companion{}   & \textit{total}   & $1481-1400$ & $238\!-\!250$   & 2.11     \\
  \enddata
  \tablenotetext{a}{Masses are calculated at the luminosity distance of the systemic optical velocity of \ugc{}.}
\end{deluxetable}\label{t:hicomponents}

The overall morphological and kinematic picture of \ugc{} as unveiled by the VLA data 
reveals a highly disturbed gas reservoir. It is striking to note that the 
\HI{} is systematically
redshifted with respect to the systemic velocity of \ugc{} and that there is 
a strong dearth of \HI{} in the center of the galaxy. Although a 
full suite of matched simulations is beyond the scope of this work, the 
asymmetry of the system in velocity space argues against a 
scenario in which the gas morphology results from purely secular effects -- 
again pointing to a tidal origin of the gas tails around \ugc{}.

Finally, we note that given the velocity difference (with respect to the stellar systemic velocity) and extent of the tidal features,
and the expected total mass of the dwarf pair, the \HI{} of \ugc{} should remain bound to \ugc{} and/or the eventual product of the merger between 
\ugc{} and \companion{}. 

\subsection{\HI{} morphology and kinematics of \companion{}}\label{s:results:companion}
In contrast to the complex morphology and kinematics of \ugc{}, the \HI{} reservoir 
of \companion{} is relatively undisturbed.
The systemic velocity of \companion{} is 1640 $\kms{}$ (see \autoref{t:neighborhoodproperties}), as measured from the SDSS spectra. 
Unlike \ugc{}, the \ion{H}{1} peak velocity is well-aligned with the 
optical emission lines of \companion{}. 

\begin{figure*}
    \centering
    \includegraphics[width=\textwidth]{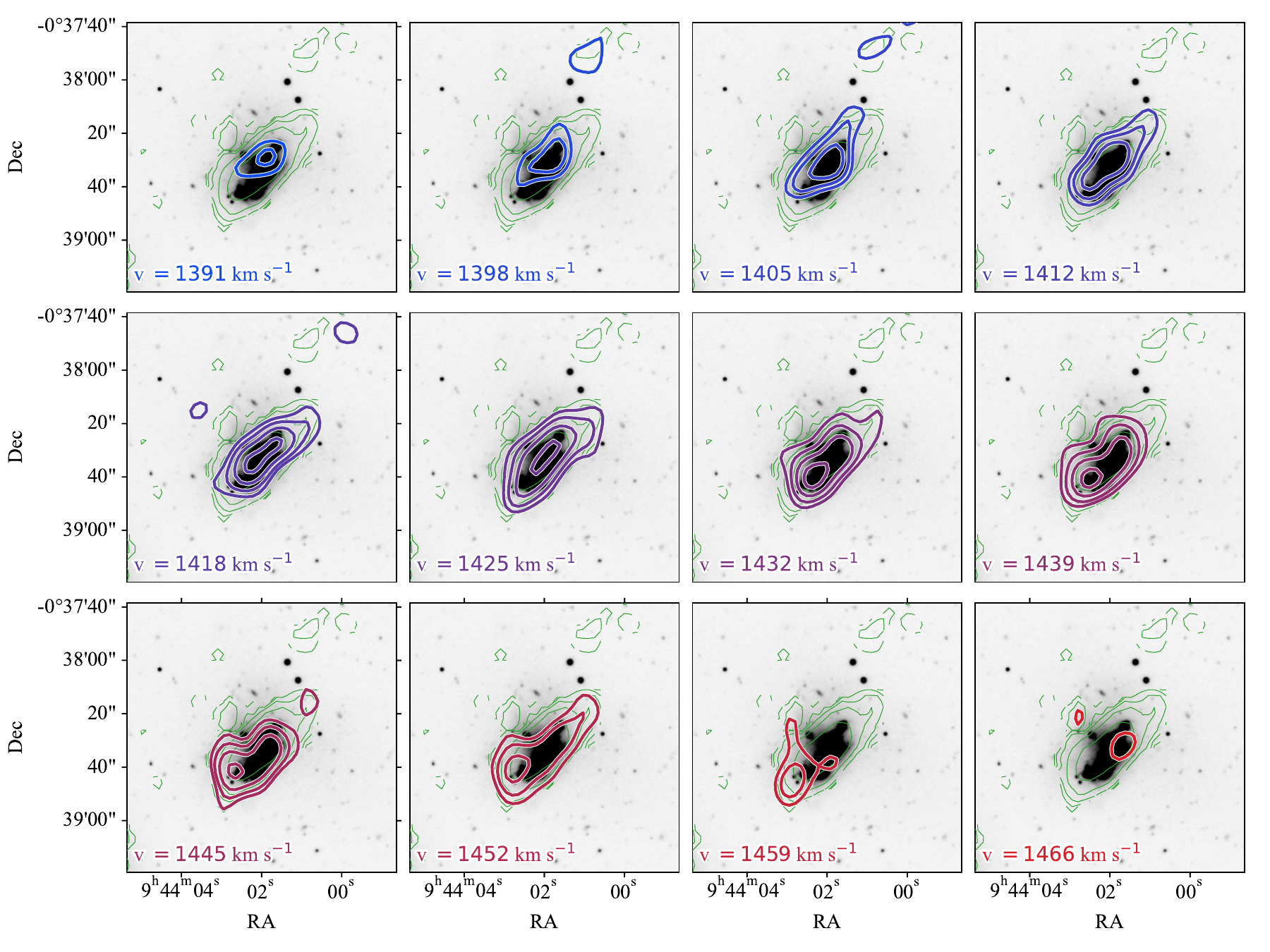}
    \vskip -5mm
    \caption{
    The same as \autoref{fig:ChanUGCa}, but for channels with significant emission around \companion{}. There is evidence for 
    disturbance in the \HI{} profile of \companion{} at 
    $1391\leq v \leq1405\kms{}$, but the bulk of the \HI{} associated with 
    \companion{} appears consistent with ordered rotation (as corroborated in  \autoref{f:barolo}).
    }
\label{fig:ChanUGCc}
\end{figure*}

In \autoref{fig:ChanUGCc} we present the channel maps for \companion. As 
is implied in \autoref{f:moment_one}, we see a gradient in velocity, with the lowest-velocity gas arising from the North-west component of the galaxy (at $\sim 1400$~km~s$^{-1}$), and the highest-velocity gas 
arising from the Southeast (at $\sim 1470$~km~s$^{-1}$). 
The channel maps of \companion{} show evidence for significant non-rotational motion,
with emission in the southeast extending over a range of $v_{\rm los}\sim70\kms{}$. 

We also see evidence for a more regular rotation curve across the face of this galaxy, though as noted above there is significant evidence for disturbance.
We estimate the
rotational velocity of \companion{} in two ways: by constructing an integrated spectrum
of the \HI{} associated with \companion{} and measuring $W_{50}$, the full width at 
half beam power, and using the 3D tilted ring fitter \bbarolo{}, which is 
well-suited to low resolution kinematic analysis \citep{diteodoro2015}, to 
fit a rotation curve (and thus an associated $v_{\rm rot}$) to the VLA cube localized around \companion{}. We fix the inclination to $i=54$ deg following the published inclination from the HyperLEDA database. 
Following \cite{ponomareva2021}, for a well-ordered rotating 
disc galaxy these two quantities should be related as $2 v_{\rm rot} \approx W_{50}$.

\begin{figure*}
    \centering
    \includegraphics[width=\linewidth]{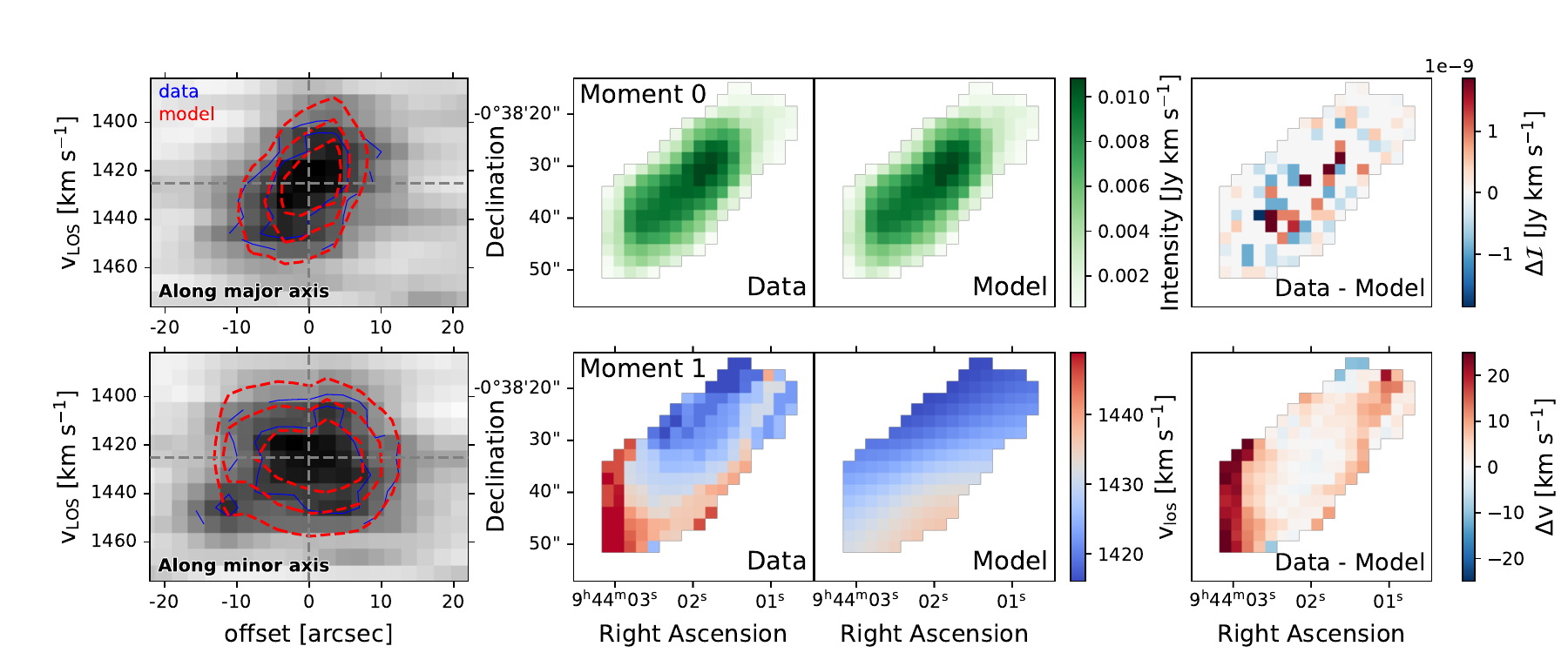}
    \caption{
        The best-fit 3D tilted ring model of \bbarolo{} for \companion{}. In the left column we show the
        PV diagram along the major (top) and minor (bottom) axes. The data are shown in greyscale with the
        model (red) and analogous data (blue) contours overlaid at the same levels. Grey dashed lines 
        mark the spatial center and systemic velocity of the system. 
        The middle set of panels show the real (left) and model (right) first (top row) and second moment maps (bottom row)  
        as calculated by \bbarolo{}. The right column shows the difference between the real and model moment
        maps. Although the tilted ring model can describe the bulk rotational
        motion in the main body of \companion{}, it does not reproduce the redshifted material in the 
        southeast region of \companion{}. We suggest that this material is the receding 
        component of a stripped gas tail that can be seen as blueshifted material to the northwest in 
        the bottom right panel of \autoref{f:moment_one}.
    }
    \label{f:barolo}
\end{figure*}

Our best-fitting \bbarolo{} model is shown in \autoref{f:barolo}. We find
an average $v_{\rm rot}=28 \pm 5 \kms{}$, though 
we find that there is significant structure to the velocity map of \companion{} that is not well-fit by the tilted ring model. This is not surprising given that there is 
significant 21cm emission by \HI{} that has been stripped from the main body of the 
galaxy, as shown in \autoref{f:moment_zero}. When we construct an integrated spectrum
containing all 21cm emission associated with \companion{}, we measure 
$W_{50}=73 \pm 3 \kms{}$, somewhat higher than the expected $W_{50}=2v_{\rm rot}$ expectation. When we exclude the emission from \HI{} that is spatially offset from the 
main body of \companion{}, we find $W_{50}=62 \pm 2 \kms{}$, in good agreement with 
the $v_{\rm rot}$ measurement from \bbarolo{}. 

Given these kinematic measures of the \HI{} associated with \companion{}, we can attempt to place the galaxy on the baryonic Tully-Fisher relation \citep{ponomareva2021}. We caution that we expect that both kinematic measures will be biased somewhat high given that the gas of 
\companion{} shows evidence of disturbance. We estimate the total baryonic mass of 
\companion{} as 
\begin{equation}
  \begin{split}
    M_{\rm bar} &= 1.4 M_{\rm HI} + M_\star\\
    &= 3.6\times 10^8 M_\odot,
  \end{split}
\end{equation} 
where the multiplicative factor of 1.4 is a correction for the mass in Helium and heavier elements. We 
ignore the mass contribution of \HH{} both  because the gas reservoir is expected to be dominated by \HI{} at these stellar masses, with \HH{} contributing $M_{H_2}/(M_{HI} + M_{H_2})\lesssim0.2$ \citep{leroy2008, kadofong2022c}, and to follow literature norms. In particular, from the measured baryonic Tully-Fisher relation (bTFR) of \cite{ponomareva2021}, which traces the correlation between total baryonic mass and 
galaxy rotational velocity,
we expect a baryonic mass of $M_{\rm bar}=1.4_{-0.9}^{+2.2}\times 10^8 M_\odot$, where the
uncertainty reflects both the uncertainty of the inferred velocity and the scatter in the 
bTFR. \companion{} therefore lies within one sigma of the 
measured bTFR, which is consistent with the interpretation that ordered rotation dominates the \HI{} velocity field.


\subsection{Integrated Star Formation Rates: H$\alpha$ and UV Estimates}\label{s:results:sfr}
Although \ugc{} shows no sign of \halpha{} emission across its face, it is well-detected in the UV. The differing timescales over which \halpha{} and UV emission trace star formation can therefore give us insight into the 
recent star formation history of \ugc{} and \companion{}.

We estimate the integrated star formation rate across the main bodies of \ugc{} and \companion{} from 
both GALEX UV photometry and the SDSS H$\alpha$ emission -- or, in the case of \ugc{}, an upper 
limit thereof. To prevent systematic offsets between the UV and \halpha{}-derived star formation 
rates, we compute the $L_{UV}$-SFR and $L_{H\alpha}$-SFR relations using the python 
implementation \textsf{pythonfsps} of the Flexible 
Stellar Population Synthesis \citep[fsps,][]{conroy2010} framework assuming the Binary Population 
and Spectral Synthesis \citep[BPASS][]{eldridge2016} stellar isochrone library and a 
Kroupa initial mass function (IMF) \citep{kroupa:2001}. We assume that the star formation rate 
has been constant for the past 10 Myr to compute $L_{H\alpha}$ and constant for the past 300 Myr 
to compute $L_{UV}$. This yields a conversion of:
\begin{subequations}
  \begin{equation}
    \frac{\text{SFR}(H\alpha)}{[M_\odot\text{ yr}^{-1}]} = 5.03\times 10^{-42}\frac{L_{H\alpha}}{[\text{erg s}^{-1}]}
  \end{equation}\label{e:sfrha}
  \begin{equation}
    \frac{\text{SFR}(FUV)}{[M_\odot\text{ yr}^{-1}]} = 1.06\times 10^{-28}\frac{L_{FUV}}{[\text{erg s}^{-1}\text{ Hz}^{-1}]}
  \end{equation}\label{e:sfrfuv}
\end{subequations}
These calibrations differ significantly from the \cite{kennicutt:1998ab} conversions as published, but 
are within $5\%$ and $25\%$ of the conversions when the difference between the chosen IMF 
is taken into account. 

We use imaging from the GALEX Medium Imaging Survey (MIS)\footnote{\rrr{The GALEX observations used in this work can be found in the Mikulski Archive for Space Telescopes (MAST) via \dataset[10.17909/qj70-dv28]{http://dx.doi.org/10.17909/qj70-dv28}.}} to measure FUV and NUV fluxes in 
the main bodies of \ugc{} and \companion{}, where we determine the geometry of the elliptical aperture\footnote{We set the limit of the aperture to $R=3$, in accordance with the approximate isophotal limit where the ellipse is defined as $\textsf{CXX}\Delta x^2 + \textsf{CYY}\Delta y^2 + \textsf{CXY}\Delta x \Delta y = R^2$ using the standard definition of the \textsf{SExtractor} geometric parameters set forth in \citep{sextractor}}
using \textsf{sep}, a python implementation of the SExtractor object segmentation algorithm \citep{sextractor}. 
Galactic extinction in both bands is corrected for using the \cite{schlafly:2011aa}
galactic reddening measurements\footnote{Obtained via the IRSA Galactic Dust Reddening and Extinction database: \dataset[10.26131/IRSA537]{http://dx.doi.org/10.26131/IRSA537}}. 

The estimation of an \halpha{}-based star formation rate is straightforward for \companion{} and 
somewhat less so for \ugc{}, given the lack of \halpha{} across the main body of the galaxy. 

We measure $\text{EW}_{H\alpha, \text{em}}$ of \companion{} from its galactic extinction-corrected 
SDSS spectrum, where we 
fit the \halpha{}, \ionline{N}{2}{6548}, and \ionline{N}{2}{6583} lines simultaneously as 
equal-width Gaussians with a scalar continuum. We additionally fit H$\beta$ in order to 
correct for internal reddening via the Balmer decrement. 

Let us now consider an upper limit for the equivalent width of the \halpha{} emission in the main 
body of \ugc{}. 
Integral Field Unit (IFU) spectroscopy from the Wide Field Spectrograph (WiFeS) did not detect \halpha{} emission across the main body of \ugc{} \citep{Pracy:2012}. The strong Balmer absorption lines in the blue imply an upper limit of $\rm EW_{H\alpha} < 5 \AA$ across the main body of \ugc{} after stellar absorption corrections, but neither emission nor absorption were detected at the expected observed wavelength of \halpha{}. We can therefore only place an upper limit on the 
\halpha{}-based SFR of \ugc{}. To be consistent with the \halpha-based SFR we will estimate for \companion{}, we use the central SDSS spectrum of \ugc{} to estimate an upper limit on 
the \halpha{} emission. As above, we fit the \ion{N}{2} lines in the vicinity of \halpha{} in addition to \halpha{}, $\rm H\beta$, $\rm H\gamma$, and $\rm H\delta$. We fix the 
equivalent width of \halpha{} to be half that of the other Balmer lines following 
\cite{gonzalezdelgado1999}. From this procedure, we estimate an upper limit of $\rm EW_{H\alpha}<3.1$ \AA{} after correction for stellar absorption and $\rm EW_{\rm H\alpha} < 0.35$ \AA{} if we ignore the absorption correction.

To convert \halpha{} equivalent widths to \halpha{} luminosities, we adapt the method\footnote{In particular, we do not include the effect of stellar absorption, as we have already taken this into account during line fitting, or of redshift, given the proximity of our target galaxies.} of \cite{bauer2013}
as follows:
\begin{equation}
  \frac{L_{H\alpha}}{[\text{erg s}^{-1}]} = 3\times10^{25}\frac{EW_{H\alpha}}{\lambda_{H\alpha, \rm obs}} 10^{\frac{M_r-34.1}{-2.5}} \left( \frac{F_{H\alpha}/F_{H\beta}}{2.86}\right)^{2.36},
\end{equation}
where our absolute r-band magnitudes are tabulated for each galaxy in \autoref{t:neighborhoodproperties}. \rrr{We note that this method assumes that the equivalent width measured in the central SDSS spectrum is equal to the average equivalent width across the face of the galaxy. Our estimate is therefore likely to be biased high given that the SDSS fiber is placed on the galaxy's central starburst (we show the placement of the SDSS fiber, as well as the optical spectrum of \companion{}, in \autoref{s:appendix:sdsspgc}) -- as such, the integrated \halpha{} SFR for \companion{} should be taken only as a rough estimate to demonstrate the galaxy's starbursting nature.}

\autoref{f:SFMS} shows a comparison of our \halpha{} and UV-based star formation rate estimates, along 
with a local comparison sample of Local Volume Legacy galaxies \citep{dale2023}. Here, the 
filled square symbols show the UV-based SFRs of \ugc{} and \companion{}, while the 
filled circle (in the case of \companion{}) and the unfilled triangle (for \ugc{}) show the \halpha{}-based SFR and upper limit, respectively. Here, one 
may immediately notice that although the UV-based SFR estimates of 
\ugc{} and \companion{} place both galaxies on the SFMS as defined by 
\cite{dale2023}, there has been a marked divergence in star formation 
behavior \rrr{between the $\sim 100$ Myr timescales probed by UV emission and the $\sim 10$ Myr timescales probed by \halpha{} emission}. While the \halpha{}-based SFR of \companion{} places the 
galaxy well above the SFMS in the vigorously star-bursting regime, the 
\halpha{} non-detection of \ugc{} indicates that star formation has 
very recently quenched in the galaxy. We will return to this divergence 
of recent star formation behavior in \autoref{s:discussion}.

\subsection{The Stellar Populations of \ugc{}}\label{s:results:stellarpops}
The ionized and neutral gas in \ugc{} show 
a coordinated story of a recently quenched star formation history.
Referring back to \autoref{fig:sdsshsc}, we see that the stellar
continuum and absorption features are clearly detected in the SDSS
spectra. Thus, the star formation history as determined from 
stellar population synthesis can reveal an independent view into the
quenching timescale of the galaxy.

We model the star formation history of UGC5205 based on the four SDSS spectra shown in \autoref{fig:sdsshsc}. We utilize Prospector \citep{johnson2017, leja2017, johnson2021}
to fit non-parametric star formation histories independently to the four spectra. Allowing for a flexible and non-parametric star-formation history is crucial to capture the very recent burst and then truncation that defines post-starburst galaxies \citep{suess2022}. 

We fix the redshift to the systemic redshift of \ugc{} as measured from the stellar absorption features 
and assume a Chabrier IMF \citep{chabrier2003}. 
We set a uniform prior on the stellar mass in each spatial region between $10^4$ and $10^8\ M_\odot$ and a uniform prior on the stellar metallicity from $ -2 < \log_{10}(Z/Z_\odot) < 0.2$. 
We use the \cite{kriek2013} dust law, allowing the power-law slope to vary between $-1$ and 0.4 and the attenuation to vary between $0<A_V<2.5$. We assume young stars are attenuated twice as much as older stars, 
and fix the shape of the IR SED following the \cite{draine2007} dust emission templates with $U_{\rm min} = 1.0$, 
$\gamma_e = 0.01$, and $q_{\rm{PAH}} = 2.0$.\footnote{\rrr{We have also run the fit using $\gamma_e=0.431$  and $U_{\rm min}=4$ based on the values quoted by \cite{draine2007} for the low-mass galaxies NGC5195 and Mrk33. We find that altering these parameters do not significantly alter our inferred SFHs.}} 
We place a uniform prior on the galaxy’s intrinsic velocity dispersion between $50 < \sigma_v < 300 \kms{}$. Following \cite{suess2022b}, we include both a spectroscopic jitter term and the \texttt{Prospector} pixel outlier model, intended to prevent bad spectral pixels from skewing the output.

The model includes three early time bins ($t_{\rm look-back} > 2$~Gyr) with fixed width, then five bins with uniform stellar mass formed but flexible time edges, such that the model can determine the start and end of star formation, then a final bin with flexible time edges and flexible stellar mass formed to capture any low-level recent star formation; \citet{suess2022} show that this model is robust to determine quenching time scales.

The derived star formation histories of each
SDSS spectrum, along with an example spectrophotometric fit, are shown in \autoref{f:wren_fits}. We remind the reader that the three off-body spectra (green, red, and purple in \autoref{f:wren_fits}) are centered on bright 
young stellar clusters along the stellar tidal tail of \ugc{}.

\begin{figure}[t]
  \centering
  \includegraphics[width=\linewidth]{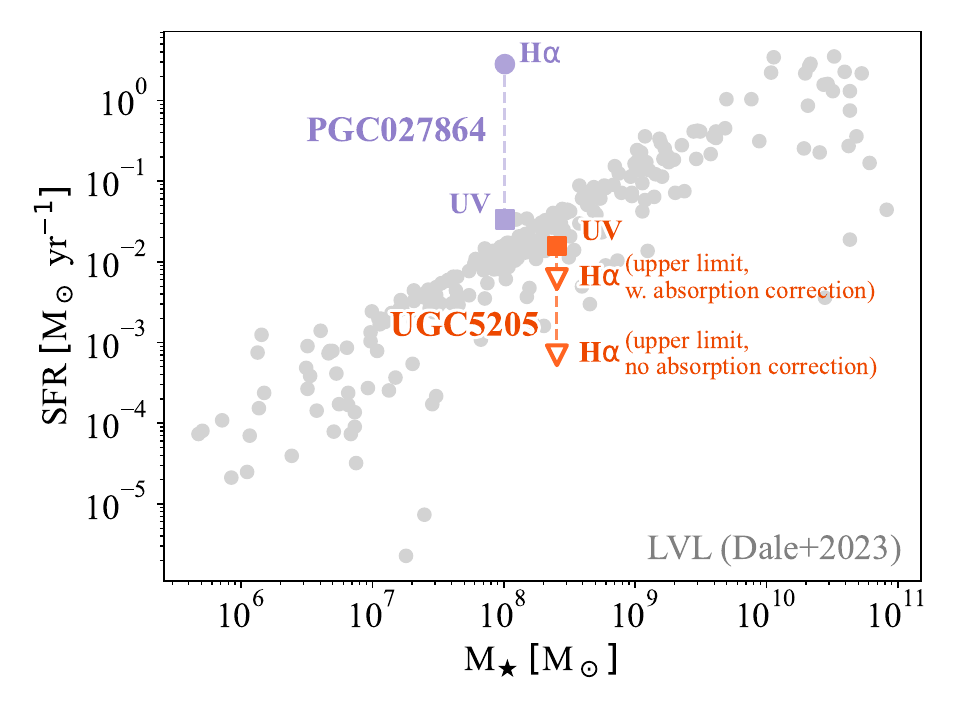}
  \caption{
    The placement of \ugc{} and \companion{} relative to the low-mass star-forming main 
    sequence (SFMS) as measured from UV-optical-IR SED fits of~\cite{dale2023}. The square symbols
    show UV-based SFR estimates, while the round circle and unfilled triangle show the 
    H$\alpha$-based SFR estimate of \companion{} and the H$\alpha$ upper limit estimate of 
    \ugc{}. We find that the UV-based SFR estimates, which reflect SFR over $\sim 100$ Myr 
    timescales, lie on or near the SFMS for both \ugc{} and \companion{}. The H$\alpha$-based 
    estimates, however, indicate strongly divergent recent ($\sim 5$ Myr) behavior of \ugc{} and  
    \companion{} wherein the \ugc{} is quenched and the SFR of \companion{} has been boosted \rrr{relative to the SFMS}
    by nearly two orders of magnitude.
  }\label{f:SFMS}
\end{figure}

\begin{figure*}[t!]
  \centering
  \includegraphics[width=\linewidth]{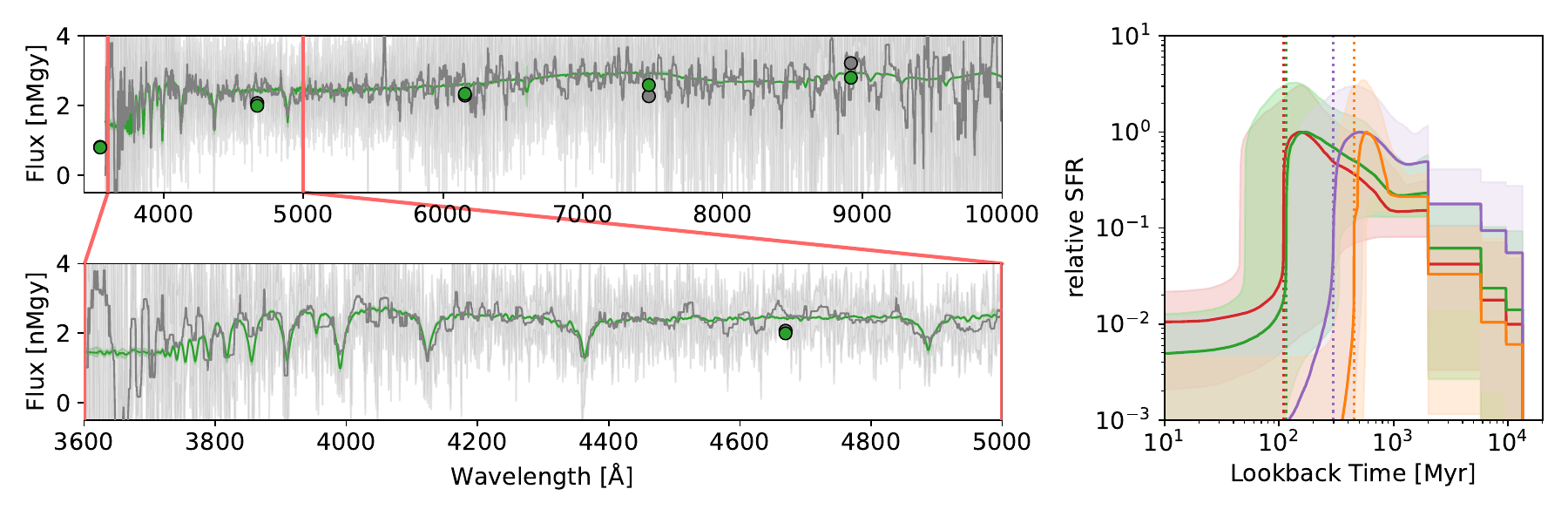}
  \caption{
      The star formation history of \ugc{} as estimated from stellar population synthesis fits 
      to the SDSS spectra and HSC-SPP photometry shown in \autoref{fig:sdsshsc}. At left we show the
      observed (grey) and model (green) spectra and photometry; the top shows the full spectrophotometric fit, while the bottom shows the
      Balmer absorption features. \rrr{The light grey curves show the original SDSS spectrum,
      while the dark grey curves show the spectrum with a median filter witha  kernel size of
      15 pixels (median size of 20 \AA{}) applied.}
      At right we show the best-fit SFH for all of the SDSS spectra taken along the tail, with the same 
      color coding as \autoref{fig:sdsshsc}. The orange curve shows the SFH derived from the central fiber, while the red, green, and purple curves show the SFH derived from the fibers placed on the stellar tidal tail. The dotted line\rrr{s} show the time at which the SFR dropped to 10\% its maximum. 
      We see evidence 
      for a starburst and quenching event between 100-300 Myr ago, consistent with the 
      stark contrast in UV- and \halpha{}-inferred star formation rates seen in \autoref{f:SFMS}. 
  }\label{f:wren_fits}
  \end{figure*}

We recover a broadly consistent star formation history from all spectra considered, though we caution against an over-interpretation of the 
seemingly earlier quenching time of the central 
spectrum (orange) due to the more complex stellar population filling the 
central fiber. The SDSS spectra are well-represented by a post-starburst 
system that underwent a coordinated burst and subsequent 
quenching event 100-300 Myr ago.
This is highly consistent with the picture of a post-starburst system presented in 
\autoref{f:SFMS}, and places a timescale for the quenching event that is independent of the 
ionized gas emission.

\subsection{Neighboring Galaxies}\label{s:results:neighbors}
By standard definitions, \ugc{} is considered a field dwarf galaxy in that it is separated by more than 2.5 Mpc and $1000\kms{}$ from the nearest massive galaxy \citep{geha_2012}. 
However, in addition to \companion{}, there are two galaxies in the vicinity of \ugc{}, \lurker{} and \bigneighbor{}. 

As shown in \autoref{t:neighborhoodproperties}, \lurker{} is close in stellar mass to \ugc{} and \companion{}. \bigneighbor{}, with an absolute 2MASS Ks-band magnitude of $M_{K_s}=-22$ \citep{twomass} and an estimated stellar mass of $1.93\times 10^{10} M_\odot$, lies just under the cutoff to be considered massive 
by the \cite{geha_2012} criterion ($M_{K_s}<-23$ or $M_\star > 2.5\times 10^{10} M_\odot$).
These galaxies are selected to be within two projected virial radii of \ugc{} at their redshift and 
within $\Delta v < 1000\kms{}$, though
both are significantly offset ($|\Delta v| \sim 300 \kms{}$) than \companion{} ($|\Delta v| \sim 50\kms{}$) from \ugc{} in line-of-sight velocity. 

\begin{figure}[t]
  \centering
  \includegraphics[width=\linewidth]{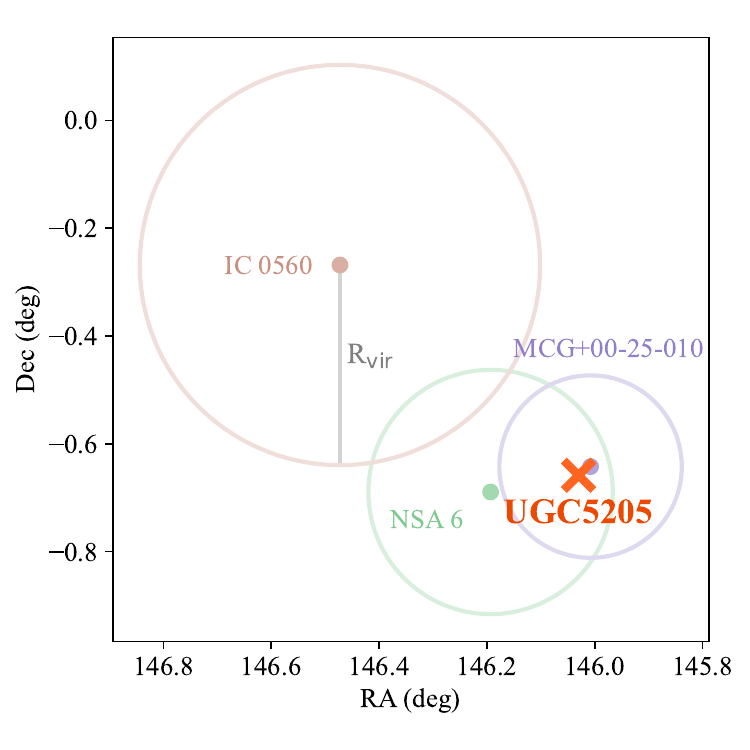}
  \caption{
    The position of \ugc{} (red cross) compared to nearby galaxies with a velocity difference of $\Delta v< 1000\kms{}$ and a projected distance of $R<2$\rvir{}. The filled points show the location of each neighboring galaxy; we show the estimated on-sky extent of the virial radius of each galaxy at its redshift by the unfilled circles of the same color.
    \ugc{} is within \rvir{} of two low-mass galaxies, the $M_r\!\approx\!  15$ \companion{} and $M_r \!\approx\! -16$ \acquaintance{}, but beyond \rvir{} of the intermediate-mass galaxy \bigneighbor{} ($M_r\!\approx\! -19$).
  }\label{f:environment}
\end{figure}

In 
\autoref{f:environment} we show the position of \ugc{} with respect to these possible neighbor galaxies, where the bold cross indicates \ugc{} and the filled points indicate the three potential neighbors. 
The large unfilled circles around each filled point indicate the on-sky size of the galaxy's 
estimated virial radius at its redshift, where we use the stellar-to-halo mass relation of 
\cite{behroozi2019} to make a crude estimate of $M_{\rm vir}$ of each galaxy and estimate its 
virial radius as
\begin{equation}
  R_{\rm vir} = \left(\frac{M_{\rm vir}}{\frac{4\pi}{3}200\rho_c(z=0)} \right)^{1/3}
\end{equation}
following \cite{bryan1999}, where $\rho_c$ is the critical density. From this initial estimate, 
we find that \ugc{} lies within the virial radius of \companion{} and \lurker{}, but outside that of 
\bigneighbor{}. This separation from \bigneighbor{} is especially important given that the 
star formation behavior of \ugc{} is unusual for a non-satellite galaxy. As detailed in \autoref{s:discussion:rampressure}, we find that
\bigneighbor{} is unable to exert sufficient ram pressure to strip the 
gas reservoir of \ugc{}.

We 
detail the optical properties of all three galaxies in the vicinity of \ugc{} (and \ugc{} itself)
in \autoref{t:neighborhoodproperties}. We retrieve redshifts and absolute r-band magnitudes for 
each galaxy from the NASA Sloan Atlas \citep{blanton:2011aa} with the exception of \ugc{}, where we 
remeasure the redshift from the absorption features in the SDSS spectrum placed on the main body of 
the galaxy. Stellar mass estimates are then made using the $g$ and $r$-band NSA photometry 
and with the color-mass relation of \cite{kadofong2022b}, which was calibrated against SED-based 
stellar mass estimates in the dwarf regime.

\begin{deluxetable}{lcccc}
  \tablecaption{Optical properties of UGC5205 and nearby galaxies}
  \tablehead{
    \colhead{Galaxy} &         \colhead{$cz$} &      \colhead{$M_r$} & \colhead{$R_{\rm proj}$} & \colhead{$M_\star$}\\
    \colhead{} & \colhead{$\kms{}$} & \colhead{} & \colhead{kpc} & \colhead{$10^8M_\odot$}
  }
  \startdata{}
  UGC5205       &  1500 & -16.24 &                  -- &                           2.54 \\
  \hline{}
  NSA 6         &  1228 & -15.27 &                 50.6 &                           1.22 \\
  \companion{} &  1449 & -16.24 &                  9.8 &                           1.02 \\
  IC 0560       &  1848 & -19.15 &                271.0 &                         193
  \enddata{}
  \tablecomments{The optical properties of \ugc{} and galaxies within two projected virial radii of \ugc{}. Redshifts and absolute r-band magnitudes are drawn from the NASA Sloan Atlas except for \ugc{}, wherein the redshift was remeasured from optical absorption features in the SDSS spectrum of the main body. The stellar mass of each galaxy was inferred from the relation measured by \cite{kadofong2022b}, which was calibrated from galaxies of this stellar mass range.}\label{t:neighborhoodproperties}
\end{deluxetable}

\section{Discussion}\label{s:discussion}
Although \ugc{} is only a single galaxy, the existence of a quenched dwarf in the field is significant to our understanding of low-mass galaxy star formation regulation due to their marked rarity at the stellar mass of \ugc{} \citep{geha_2012}. In this discussion, we will consider both how \ugc{} came to be quenched and what implications the existence of a quenched field dwarf has on our understanding of star formation regulation in isolated dwarf galaxies.

\subsection{The Quenching Timescale of \ugc{}}\label{s:discussion:timescale}
The combination of a lack of \halpha{} emission, a strong UV detection (\autoref{s:results:sfr}), and a predominantly young 
stellar population (\autoref{s:results:stellarpops}) paint a consistent story of a galaxy that underwent a global quenching event $\sim\!100\!-\!300$ Myr ago.

We can also make a rough, back of the envelope calculation of the lookback time at which the tidal features
seen in the \HI{} of \ugc{} were launched based on their extent and velocity relative to the 
systemic velocity of \ugc{}. Due to the complex dynamics of the dwarf interaction, a hydrodynamic simulation 
would be necessary to truly place a timescale on the merger interaction, but here we can at least 
make a rough consistency check of the merger-driven quenching picture.

The \HI{} tails B1 and B2 are roughly 56\arcsec{} and 72\arcsec{} in extent, which corresponds to 
5.8 kpc and 7.5 kpc at the distance of \ugc{} \rrr{assuming an angular diameter distance of $d_A=21.3$ Mpc based on its recessional velocity of $1500\kms{}$}. 
Their mean velocities, as determined by a geometric mean 
of the moment one map, are $1520\kms{}$ and $1544\kms{}$, respectively. If we simply estimate the creation 
timescale assuming a constant velocity, we arrive at a launching lookback time of $t_{lb}=280$ Myr and 
$t_{lb}=160$ Myr for B1 and B2 \rrr{assuming that the transverse velocity is equal to the difference in velocity between the tidal tail and the systemic stellar velocity of \ugc{}}. This is consistent with the picture wherein the tidal features were 
pulled out of the \HI{} reservoir during or shortly after the starburst event, and strengthens the 
interpretation of a interaction-driven quenching picture for \ugc{}.

\subsection{HI Properties} 
Prior to the resolved \HI{} data presented in this work, single dish observations revealed a 
large reservoir of cold gas associated with \ugc{} \citep{springob2005}. 
It was unclear, then, how \ugc{} could support a large reservoir of \HI{} and yet be completely devoid 
of recent star formation.

The spatially-resolved VLA observations presented in this work show that the bulk of the \HI{} 
reservoir of \ugc{} is indeed not spatially coincident with the bulk of the stellar mass of the galaxy
and has instead been pulled out into two large tidal tails. Furthermore, both tails are 
offset in velocity from the systemic redshift as measured from stellar absorption features in the same direction. Indeed, the \HI{} morphology of 
\ugc{} suggests that the lack of recent star formation in the galaxy is borne out of a dearth of cold gas in the galaxy center and not an anomalously low star formation efficiency, as may be suggested from single dish observations alone.

Along these lines, we find that the \HI{} morphology of \companion{} -- the starbursting neighbor of \ugc{} -- is relatively undisturbed. With the majority of its cold gas remaining near its center, \companion{} has much more cold gas available for star formation than \ugc{} despite a similar total \HI{} mass. 
The difference in the spatial distribution of the cold gas provides a clear explanation for the divergent star formation behavior of \ugc{} and \companion{}. 
Furthermore, the divergence of morphological and kinematic properties of this dwarf pair -- despite their similar global \HI{} properties -- underlines the need for interferometric imaging to shed light on the physical processes that shape galaxies' gas reservoirs.




{}

\subsection{Implications for Quenching Dwarfs in the Field}
It has been shown that the build-up of a quiescent galaxy population at $M_\star \lesssim 10^{8.5} M_\odot$ is 
negligible for non-satellite galaxies. The properties of \ugc{} demonstrate that dwarfs can indeed quench outside of the
influence of a massive host for short timescales ($5\!\lesssim\!\tau\!\lesssim \!500$ Myr), while the dearth of a 
larger population of similar galaxies implies that \ugc{} will not remain quenched. 

The spatial distribution of the \HI{} associated with \ugc{} implies that the galaxy has entered a period of quiescence through its interaction with the nearby dwarf \companion{}. In this section, we will explore the 
implication of \ugc{}'s quenched status on the overall occurrence rate of 
quenched field galaxies.

{}

\subsubsection{Dwarf-Dwarf Mergers}
Both projected distance and velocity difference point to a clear ongoing interaction between \ugc{} and \companion{}. Indeed, \companion{} is currently undergoing a strong starburst; such starbursting periods have been shown to correlate
with major mergers between low-mass galaxies
\citep{stierwalt2015,privon2017,kado-fong:2020aa}. As shown in \autoref{fig:integratedspectra} and \autoref{fig:ChanUGCc}, however, there is little evidence to indicate a disturbance in the \HI{} reservoir of \companion{}. This asymmetry in outcomes implies an asymmetry in the pair which allows \companion{} 
to retain its \HI{} reservoir while presumably instigating significant tidal disruption in \ugc{}, despite their similar stellar masses (see \autoref{t:neighborhoodproperties}). Potential asymmetries include highly different 
mass-to-light ratios (though we note that we have demonstrated that \companion{} lies on the baryonic Tully-Fisher Relation) or a difference in the concentration of the original
\HI{} profile which made \ugc{} relatively more susceptible to stripping.

Regardless of the reason for the asymmetry, however, the existence of 
\ugc{} shows that interactions between low-mass galaxies may provide a 
viable way to quench star formation in field dwarfs. The tidal morphology and kinematics of the \HI{} around \ugc{} as well as the coincident starburst and tidal feature creation paint a consistent 
picture in which the interaction between \ugc{} and \companion{} both triggered star formation in 
\companion{} and quenched star formation in \ugc{}.

{}

Finally, as briefly noted in \autoref{s:results:ugc}, the gaseous tidal features around \ugc{} remain 
bound to the galaxy. This means that the cold gas associated with \ugc{} 
should eventually reaccrete onto the system and reignite star formation. In particular, using the approximate halo properties from \autoref{s:results:neighbors} and the halo mass-concentration relation of \cite{child2018}, we estimate a freefall time of 
\begin{equation}
    \begin{split}
        t_{ff} &= \sqrt{\frac{3\pi}{32 G\rho}}\\
        &\approx 260{\ \rm Myr}
    \end{split}
\end{equation}
at the largest point of extent of the \HI{} tidal tails. From our inferred SFH of \autoref{s:results:stellarpops}, \ugc{} has already been quenched for $\lesssim\!300$ Myr, implying that the total time that \ugc{} will be quenched by this interaction is $t\lesssim 560$ Myr.
We therefore suggest that 
low-mass interactions such as the system in this work provide an avenue to \textit{temporarily} quench 
star formation. Such a mechanism is in good agreement with observations that quenched dwarfs account for, at most, a few percent of low-mass galaxies in the field \citep{geha_2012}.

\subsubsection{Low-Mass Group/Filament}
Interestingly, \ugc{} is located around 2 Mpc from another quenched field
dwarf of stellar mass $M_\star\sim 2\times10^6 M_\odot$, COSMOS-dw1 \citep{polzin:2021aa}. The projected distance places 
these two galaxies a factor of several beyond each other's virial radii given their velocities, 
making a direct interaction 
between the dwarfs highly unlikely. 
However, the relative proximity of two independently discovered highly rare objects suggests that a larger scale environmental factor 
may be at play in producing these quenched field dwarfs.

\cite{polzin:2021aa} suggested internal feedback from star formation to be the mechanism which has temporarily quenched COSMOS-dw1. This interpretation is more feasible for COSMOS-dw1 given that it is a factor of $\sim 100$ lower in stellar mass than \ugc{}, but does not explain why 
two quenched dwarfs would be found within a few Mpc of each other. Similarly, our interpretation that the interaction with \companion{} has temporarily quenched \ugc{} connects more directly to the environment of \ugc{} but does not directly imply the existence of another quenched dwarf at a separation larger than several virial radii.

The moderate spatial proximity of these two quenched systems is especially puzzling given that \ugc{} has been 
identified as a void galaxy \citep{pustilnik:2019}. 
This makes it difficult to 
envision a scenario in which a large-scale cosmic filament would be responsible for
quenching both \ugc{} and COSMOS-dw1. However, we note that the void catalog of \cite{pustilnik:2019} was constructed using only those galaxies 
brighter than $M_K=-22$; it is thus possible that \ugc{} and COSMOS-dw1 are connected by a large-scale structure traced only by low-mass galaxies. 
At this point, an attempt to define such a structure based on the existence of \ugc{} and COSMOS-dw1 would be highly speculative.
A further spectroscopic exploration of other low-mass galaxies in the vicinity of these two quenched field dwarfs will be necessary to 
more conclusively establish a link between the unusual quenched nature of these two galaxies.


\section{Conclusions}

In this work we have presented new spatially resolved \HI{} observations of the interacting field
dwarf system \ugc{} and \companion{}. Despite the fact that they are remarkably similar in their integrated UV, optical, and \HI{} properties,
one dwarf is quenched ($\text{EW}_{\rm H\alpha}<2$ \AA) and the other is starbursting ($\text{EW}_{\rm H\alpha}>1000$ \AA) in this peculiar pair. The 
existence of \ugc{}, a quenched dwarf in the field, is intriguing due to the fact that the quiescent field dwarf population is still poorly understood at this stellar mass. Previous works have put an upper limit on the quenched fraction at the stellar mass of \ugc{}, but this galaxy represents a novel opportunity to directly understand the nature of quenching processes at this mass scale. 

With our new VLA observations, we have shown that:
\begin{itemize}
  \item \rrr{Approximately half of the} \HI{} reservoir of \ugc{} has been pulled into large \rrr{tidal} tails that are spatially and kinematically offset from the main stellar body of \ugc{}, making most of the cold gas in the galaxy unavailable for star formation despite \rrr{a} sizable total \HI{} mass.
  \item The \HI{} reservoir of \companion{}, conversely, shows only minor disturbance. The core can be fit by a tilted ring model, with some emission from tidal \HI{} visible along the axis of the dwarf pair.
\end{itemize}
This asymmetry in \HI{} morphology provides a clear explanation for the marked divergence in the apparent global star formation efficiency (defined simply as $\text{SFR}/M_{\rm HI}$) of \ugc{} and \companion{}. 

With this morphological information in hand, and given the lack of other likely disrupting bodies, we suggest that it is the ongoing interaction with \companion{} \rrr{that} has temporarily quenched \ugc{}. 
This interpretation is supported by the fact that the timing of the quenching event in \ugc{}, as inferred
both by the UV and \halpha{} emission (or lack thereof, \autoref{s:results:sfr}) and from stellar population
fitting of \ugc{} (\autoref{s:results:stellarpops}), is coincident with the rough lookback time at 
which the gaseous tidal features were created (\autoref{s:discussion:timescale})
Under this assertion, we can examine the implications of \ugc{} on the build-up of a population of quenched dwarfs:
\begin{itemize}
    \item \ugc{} and \companion{} show that dwarf-dwarf interactions are not only able to trigger starbursts, but also quench star formation. Both phenomena can moreover happen simultaneously in the same interaction.
    \item Given that the \HI{} in \ugc{} likely remains bound to the system, the sizable cold gas reservoir should eventually reaccrete and reignite star formation in the system. Based off our inferred SFH and an estimate of the freefall time at the position of the tidal features, we estimate a total quenched period of $\approx 560$ Myr. This suggests that interaction-driven quenching events are temporary in low-mass systems.
    \item \ugc{} is 2 Mpc away from the lower mass quenched field dwarf COSMOS-dw1. Given their low masses, this distance is too far for a direct interaction between the quenched dwarfs. Furthermore, while \ugc{} is highly consistent with a pair interaction quenching picture, COSMOS-dw1 has been suggested to have self-quenched through star formation feedback. However, it is possible that a large-scale structure populated by low-mass objects could increase the efficacy of both quenching mechanisms.
\end{itemize}
The dwarf pair demonstrates that the star formation behavior of low-mass galaxies is sensitive not only to the presence of massive host galaxies, as has been studied in depth in the literature \citep[see, e.g.][]{geha_2012, geha_2017, carlsten:2020survey, carlsten:2021structure, mao_2021, carlsten:2022survey}, but also to the presence of other low-mass galaxies, which has previously been explored only in the enhanced star formation scenario \citep[see, e.g.][]{stierwalt2015, pearson2016, privon2017, kado-fong:2020aa, subramanian2023}. 

The existence of \ugc{} and its low-mass neighbors demonstrate the potential of a new avenue for understanding star formation regulation in dwarf galaxies outside the influence of a massive host.    
Characterizing just one of these rare quiescent field dwarfs has shed new light on the role of low-mass interactions in 
regulating dwarf star formation -- assembling a larger sample of quenched field dwarfs would provide even more novel constraints on the duty cycle and efficiency of such quenching events. 
\tinytitans{} has established a sample of dwarf multiples for galaxies with SDSS spectroscopy; the upcoming generation of deeper surveys will, on timescales within a few years, push this pair-finding out to larger volumes and lower masses \citep{darraghford2022, luo2023} and shed new light into the regulation and evolution of star formation in low-mass galaxies.

\vskip 5.8mm plus 1mm minus 1mm 
The authors thank George Privon for helpful conversations that improved the quality of this manuscript. \rrr{The authors also thank the anonymous referee for their insightful comments that improved the quality of this work.}

EKF gratefully acknowledges support from the Yale Center for Astronomy and Astrophysics Prize Postdoctoral Fellowship. 

The National Radio Astronomy Observatory is a facility of the National Science Foundation operated under cooperative agreement by Associated Universities, Inc.

Basic research in radio astronomy at the U.S. Naval Research Laboratory is supported by 6.1 Base Funding. 

The Hyper Suprime-Cam (HSC) collaboration includes the astronomical communities of Japan and Taiwan, and Princeton University. The HSC instrumentation and software were developed by the National Astronomical Observatory of Japan (NAOJ), the Kavli Institute for the Physics and Mathematics of the Universe (Kavli IPMU), the University of Tokyo, the High Energy Accelerator Research Organization (KEK), the Academia 
Sinica Institute for Astronomy and Astrophysics in Taiwan (ASIAA), and Princeton University. Funding was contributed by the FIRST program from the Japanese Cabinet Office, the Ministry of Education, Culture, Sports, Science and Technology (MEXT), the Japan Society for the Promotion of Science (JSPS), Japan Science and Technology Agency (JST), the Toray Science Foundation, NAOJ, Kavli IPMU, KEK, ASIAA, and Princeton University. 

This paper makes use of software developed for Vera C. Rubin Observatory. We thank the Rubin Observatory for making their code available as free software at http://pipelines.lsst.io/.

This paper is based on data collected at the Subaru Telescope and retrieved from the HSC data archive system, which is operated by the Subaru Telescope and Astronomy Data Center (ADC) at NAOJ. Data analysis was in part carried out with the cooperation of Center for Computational 
Astrophysics (CfCA), NAOJ. We are honored and grateful for the opportunity of observing the Universe from Maunakea, which has the cultural, historical and natural significance in Hawaii. 

Based on observations made with the NASA Galaxy Evolution Explorer. GALEX is operated for NASA by the California Institute of Technology under NASA contract NAS5-98034.  

Funding for the SDSS and SDSS-II has been provided by the Alfred P. Sloan Foundation, the Participating Institutions, the National Science Foundation, the U.S. Department of Energy, the National Aeronautics and Space Administration, the Japanese Monbukagakusho, the Max Planck 
Society, and the Higher Education Funding Council for England. The SDSS Web Site is http://www.sdss.org/. The SDSS is managed by the Astrophysical Research Consortium for the Participating Institutions. The Participating Institutions are the American Museum of Natural 
History, Astrophysical Institute Potsdam, University of Basel, University of Cambridge, Case Western Reserve University, University of Chicago, Drexel University, Fermilab, the Institute for Advanced Study, the Japan Participation Group, Johns Hopkins University, the Joint 
Institute for Nuclear Astrophysics, the Kavli Institute for Particle Astrophysics and Cosmology, the Korean Scientist Group, the Chinese Academy of Sciences (LAMOST), Los Alamos National Laboratory, the Max-Planck-Institute for Astronomy (MPIA), the Max-Planck-Institute for 
Astrophysics (MPA), New Mexico State University, Ohio State University, University of Pittsburgh, University of Portsmouth, Princeton University, the United States Naval Observatory, and the University of Washington.






\bibliography{dwarf}

\appendix{
    \section{\rrr{SDSS Spectrum of \companion{}}}\label{s:appendix:sdsspgc}
    \rrr{In addition to the four spectra taken of \ugc{}, there is one SDSS spectrum 
    associated with the companion \companion{}. }q      
    \rrr{In \autoref{fig:sdsspgc}, we show this spectrum and its placement with respect to the galaxy; unlike \ugc{}, the optical spectrum shows strong line emission that indicates that \companion{} is actively starbursting. }
    
    \rrr{As in the analogous main text figure (\autoref{fig:sdsshsc}), at left we show a $gri$-composite image of the \ugc{} and \companion{} system with the position of the SDSS fiber on \companion{} overlaid. The inner circle shows the on-sky size of the 3'' diameter fiber. At right we show the spectrum itself; unlike \autoref{fig:sdsshsc}, we do not show a binned version of the spectrum due to the high SNR of the emission lines.}

    \begin{figure*}[b]
      \centering     
      \makebox[\textwidth][c]{\includegraphics[width=1.2\textwidth]{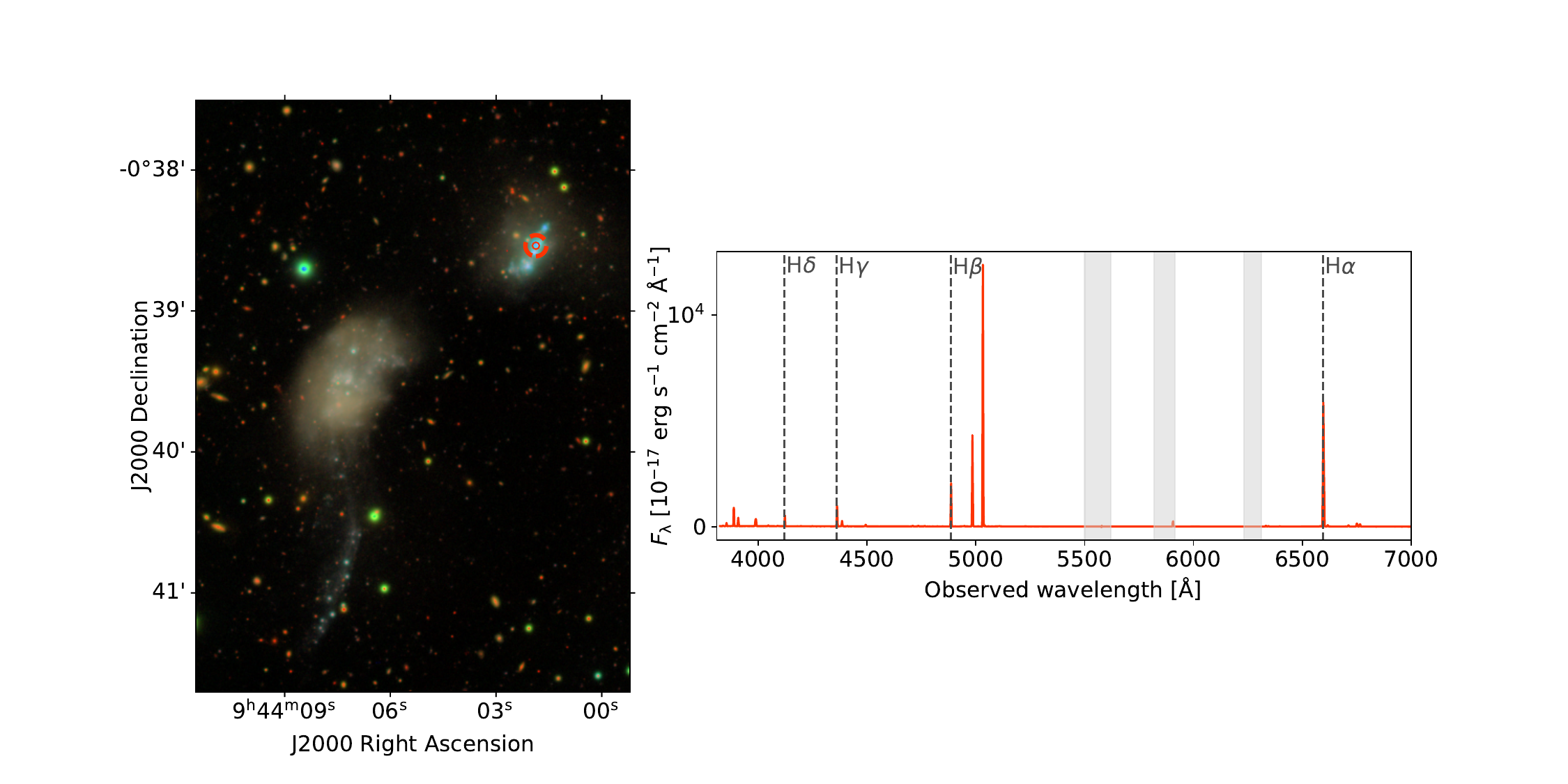}}
      \vspace{-25pt}
      \caption{\rrr{The same as \autoref{fig:sdsshsc}, but showing the SDSS spectra associated with the companion \companion{}. Unlike \ugc{}, strong line emission is clearly present in the central spectrum of \companion{}.}}
      \label{fig:sdsspgc}
    \end{figure*}    

  \section{Ancillary \HI{} data}

  \begin{figure*}[t]
  \centering
  \includegraphics[width=\linewidth]{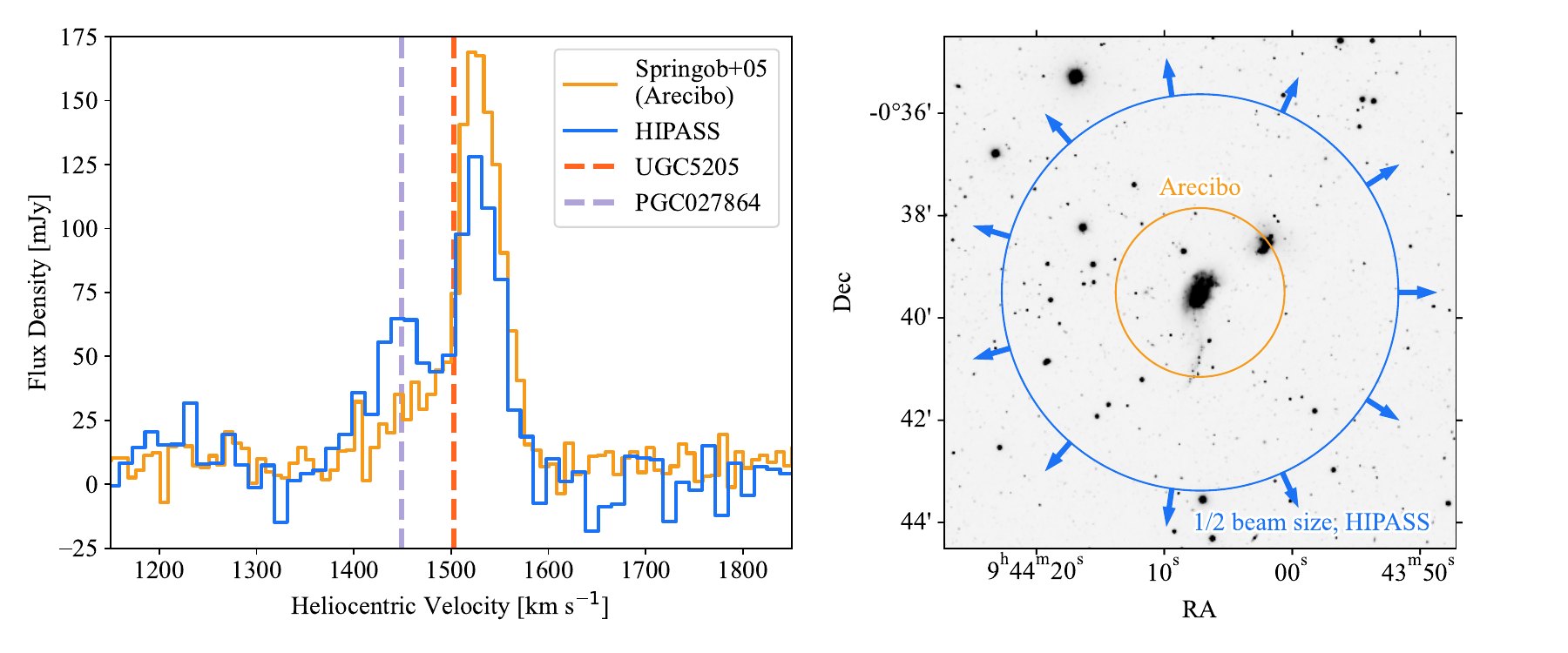}
  \caption{
    The 21cm-line spectra (left) and approximate spatial extent (right) of the archival \HI{} 
    observations of HIPASS and \cite{springob2005} centered on \ugc{}. At left, we also overlay the
    mean heliocentric velocity of \ugc{} and \companion{} as dashed orange and purple vertical lines, 
    respectively. The Arecibo beam (HPBW$=3.3$ arcmin) covers all of \ugc{} and part of \companion{}, as shown on the right. The HIPASS beam (gridded beamsize of 15.5 arcmin) covers both galaxies; in the 
    right panel, the HIPASS beam extends beyond the figure extent and so for illustrative purposes 
    we plot the aperture equivalent to half the HIPASS beam. 
  }\label{f:archival}
  \end{figure*}

  \subsection{Archival \HI{} Observations}\label{s:appendix:archival}
  Single dish observations of \ugc{} and \companion{} find a gas-rich system. \autoref{f:archival} shows HI line profiles from the HI Parkes All Sky Survey \citep[HIPASS,][]{koribalski2004} and Arecibo \citep{springob2005}. The HIPASS beam contains both \ugc{} and \companion{}, while the Arecibo beam is centered on \ugc{} and grazes the southwestern edge of \companion{}. The HIPASS HI profile shows a peak at the redshift of \companion{}; this gas is likely the 
  origin of the asymmetric tail in the Springob et al. (2005) line profile, as the Arecibo beam includes a part of \companion{}. The main peak of the \ion{H}{1} profiles, however, is shifted redwards of \ugc{}, as is seen in the spatially resolved VLA data.

  \subsection{Integrated Spectra}\label{s:appendix:integrated}
  In order to understand whether we are missing significant flux at spatial scales 
  larger than what can be probed by the high-resolution C-array data, the system was also observed in the D configuration. We show the moment zero map of the D-array data in \autoref{f:moment_zero_DARRAY}. Much of the
  spatial structure of the system's \HI{} (and in particular the disturbed \HI{} associated both 
  with \ugc{} and with \companion{}) is not observed in the D-array data
  due to the larger beamsize.
  
  To understand whether there is significant 21 cm emission from \HI{} at 
  larger spatial scales than can be reached by the C configuration data, we 
  integrate the spectrum coincident with \ugc{} in both the C-array and D-array
  data.
  We extract all the flux 
  from within the black circle shown in \autoref{fig:integratedspectra}. The black circle is centered on (09:44:07.714, -03.39.54.389) in order to encompass all of the emission associated with \ugc{} while excluding emission from its nearby companion \companion{}. We find a very good agreement in total flux $F$ between the two integrated spectra, with a fractional difference of $\Delta F/F=0.26$ for all emission at $\Delta F/F = 0.10$ for gas at $v>1500 \kms{}$. There is an excess at velocities of $1400$ to $1500 \kms{}$; this is at the velocity of \companion{}, and is likely due to some  gas associated with the companion contaminating the low resolution D-array data, but not higher resolution the C-array data. 

  \begin{figure}[t]
  \centering
  \includegraphics[width=0.4\linewidth]{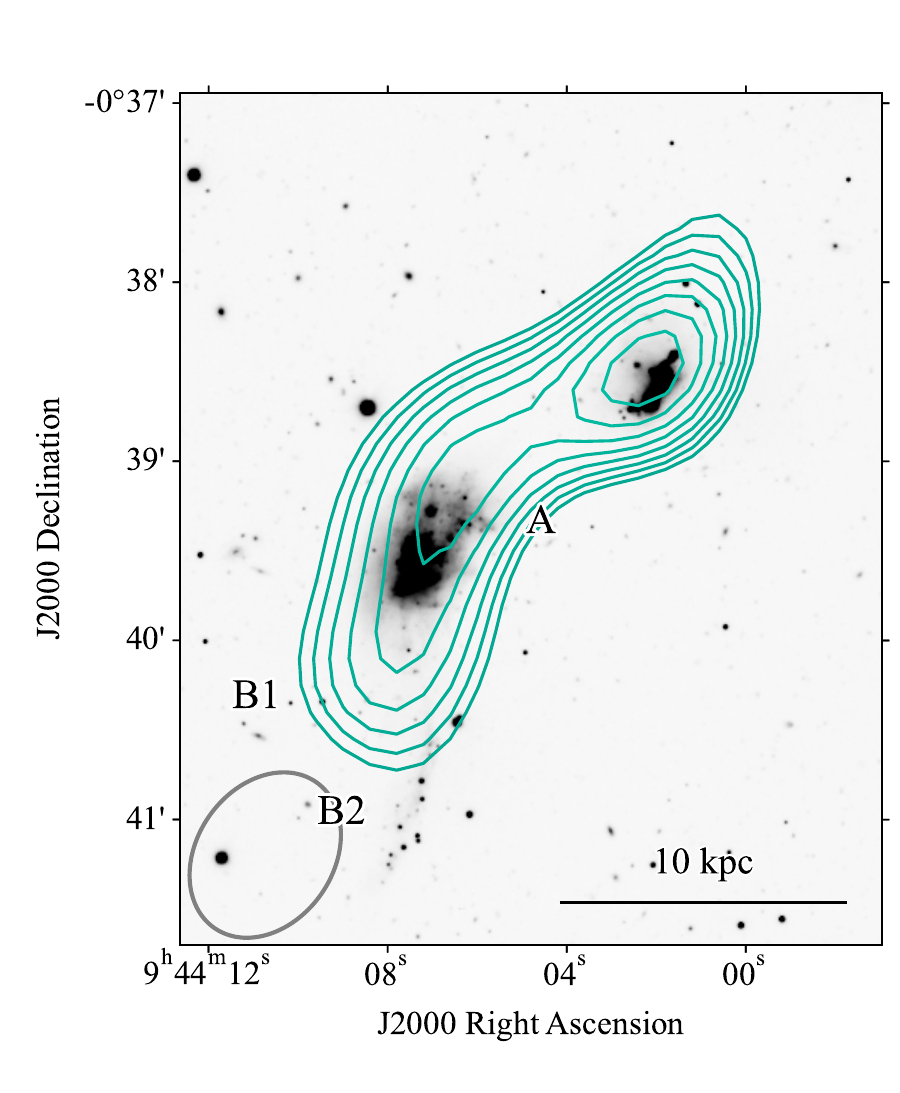}
  \caption{
    The same as \autoref{f:moment_zero} for the moment 0 map of the D-array. Again, the VLA
    data (teal contours) overlaid over HSC-SSP g-band imaging. The average beamsize of 
    the cube over which the moment 0 map was computed is shown in the bottom left of each panel.
  }\label{f:moment_zero_DARRAY}
  \end{figure}


  \section{The Plausibility of Ram Pressure Stripping by \bigneighbor{}}\label{s:discussion:rampressure}
    Given the singular nature of its current star formation and \HI{} reservoir, 
  we will consider the potential influence of the nearby intermediate-mass galaxy \bigneighbor{} on the structure and evolution of \ugc{}.

  
  
  
  Although \ugc{} is considered a field galaxy by standard methods to determine low-mass galaxy environment \citep{geha_2012}, the intermediate-mass galaxy 
  \bigneighbor{} 
  is at a velocity difference of $\Delta v = 550\ \kms{}$. We should thus examine whether this more massive galaxy is able to exert sufficient ram pressure to strip the atomic gas from \ugc{}. 
  
  Following \cite{Gunn:1972aa}, we define the ram pressure at the position of \ugc{} as
  \begin{equation}
    P_e = \rho(R_t) \Delta v^2,
  \end{equation}
  where $\rho(r)$ is the (radial) density profile of the circumgalactic medium of \bigneighbor{}. $R_t$ here is the radial distance of \ugc{} where the center of \bigneighbor{} is $R=0$. 
  The velocity difference $\Delta v=550\ \kms{}$ here refers to the velocity difference between \ugc{} and \bigneighbor{} assuming that both galaxies are at the distance of \bigneighbor{}. 
  
  Assuming plane-parallel geometry, the confining weight that the gas feels can be written as:
  \begin{equation}
    \mathcal{W}=2 \pi G \Sigma_\star \Sigma_g,
  \end{equation}
  where $\Sigma_\star$ and $\Sigma_g$ are the stellar and gas surface densities, respectively. The original treatment of \cite{Gunn:1972aa} implicitly assumes that the contribution to the weight from dark matter is negligible compared to that from stars -- while this may not be true in detail for \ugc{} due to its lower stellar mass, including the effect of dark matter would make the gas even harder to strip than in this simple exercise.
  
  For the hot gas associated with \bigneighbor{} to successfully strip the gas from \ugc{}, the ram pressure must overcome the external weight felt by the gas. Taking the NASA Sloan Atlas (NSA) photometry of \ugc{}, we arrive at an average stellar mass surface 
  density of $\Sigma_\star = 40 M_\odot \rm\ kpc\per{2}$. The relevant gas surface density differs from the present-day \HI{} surface density for two reasons. First, the relevant distribution of the gas reservoir is the distribution \textit{before} the galaxy's reservoir was disturbed. Second, our \HI{} observations account for only one phase of the galaxy's interstellar medium, while this exercise concerns the total gas surface density.
  
  In order to estimate the total, pre-disturbed gas surface density, let us assume that the \HI{} of \ugc{} was originally distributed with the same profile as the present-day stars, but with a scale radius equal to twice that of the half-light radius ($R_{1/2,\star}=1.9$ kpc). We will neglect the dynamical contribution of other phases of the ISM (notably neglecting molecular hydrogen) both because \HH{} is likely to play only a minor dynamical role in low-mass galaxies \citep{leroy2008,kadofong2022b} and because neglecting the contribution of \HH{} will lead to an underestimate in the \bigneighbor{} gas density needed to strip the gas from \ugc{} (i.e., we will over-estimate the capacity for \bigneighbor{} to strip \ugc{}). 
  
  These assumptions lead to average estimated gas surface density of $\Sigma_g=27\rm \ M_\odot\ pc^{-2}$. We thus find that for stripping to occur the density of the gas associated with \bigneighbor{} at the position of \ugc{} must exceed
  \begin{equation}
    \begin{split}
      n_H(R_t) &> \frac{2 \pi G \Sigma_\star \Sigma_g}{m_H \Delta v^2}\\
        &> 0.004\ \rm cm^{-3}.\\
    \end{split}
  \end{equation}
  
  Although simulations predict that the circumgalactic medium (CGM) of sub-L* galaxies can exceed this requirement, such densities are only predicted very nearby the galaxy \citep[$R\lesssim 0.1$\rvir{},][]{oppenheimer2016}. 
  At the present-day distance of \ugc{}, CGM densities of sub-L* galaxies are several orders of magnitude below this 
  stripping threshold ($n_H\sim 10^{-5}$ cm$\per{3}$ at $R=$\rvir{}).
  We thus conclude that the disturbed nature of the \HI{} of \ugc{} 
  is not due to ram pressure stripping from gas associated with \bigneighbor{}.
}
\end{document}